\documentclass[11pt,a4paper,epsfig]{article}

\usepackage{amsfonts}
\usepackage{epsfig}

\title{\textbf{Composite solitary waves in three-component scalar field
theory: I. The kink variety.}}
\author{A. Alonso Izquierdo$^{(a)}$ and J. Mateos Guilarte$^{(b)}$
\\ {\normalsize {\it $^{(a)}$ Departamento de Matematica
Aplicada}, {\it Universidad de Salamanca, SPAIN}}\\{\normalsize
{\it $^{(b)}$ Departamento de Fisica Fundamental and IUFFyM}, {\it
Universidad de Salamanca, SPAIN}}}
\date{}

\begin{document}

\maketitle

\begin{abstract}
We study the structure of the manifold of solitary waves in a
particular three-component scalar field theoretical model in
two-dimensional Minkowski space. These solitary waves involve one,
two, three, four, six or seven lumps of energy.
\end{abstract}

\section{Introduction}

The search for solitary waves is a mandatory topic in a huge number
of branches of non-linear science. The reason for this lies in the
fact that the method of linearizing nonlinear differential equations
omits phenomena of essential importance in many different physical
problems. The non-linear Klein-Gordon system of coupled PDE
equations, generalized to $n$-real scalar fields, reads:
\begin{equation}
\frac{\partial^2 \phi_a}{\partial t^2}-\nabla^2 \phi_a
+\frac{\partial V}{\partial \phi_a}=0 \hspace{1cm} a=1,2, \dots , N
\qquad . \label{eq:nlkg}
\end{equation}
Here, $V(\phi_1,\dots,\phi_n)$ is a differentiable potential
function of the scalar fields $\phi_a$, whereas $\frac{\partial
V}{\partial \phi_a}$ are non-linear functions of the fields. It is
clear that this system is invariant under Poincare transformations.
For this reason we shall use a system of units where the speed of
light is set to one: $c=1$. Also, due to the potential applications
of the PDE system (\ref{eq:nlkg}) in quantum physics we shall
implicitly understand $\hbar=1$, i.e., we shall work in the natural
system of units.

We shall focus on physical systems governed by the PDE system
(\ref{eq:nlkg}) in only one spatial dimension. The paradigm with
only one real scalar field is the celebrated sine-Gordon equation:
\[
\frac{\partial^2 \phi}{\partial t^2}-\frac{\partial^2 \phi}{\partial
x^2}+\sin \phi=0
\]
Arising in differential geometry in the theory of negative curvature
surfaces, the sine-Gordon equation is ubiquitous in the description
of one-dimensional physical phenomena. To mention a few, 1) it is
the evolution equation for the amplitude of various slowly varying
waves, 2) it describes the propagation of a dislocation in a
crystal, 3) it provides a model for elementary particles moving on a
line, 4) it rules the propagation of magnetic flux in a long
Josephson-junction transmission line, 5) it determines the
modulation of a weakly unstable baroclinic wave packet in a
two-layer fluid, etcetera, see, e.g., Reference \cite{Drazin}.

Point 3) deserves special mention and will be the arena within we
shall analyze the generalized non-linear Klein-Gordon PDE system
proposed in this paper. The similarities between the properties of
solitary waves (or solitons) and elementary particles are clear. The
field energy density of a non-linear wave (solitary wave or soliton)
is localized in a lump. Moreover, being non-dispersive, solitary
waves (or solitons) propagate without changes in shape. After
collisions, solitary waves almost keep their structure, leaving some
radiation in the form of dispersive waves as remnants. Solitons,
however, only suffer phase shifts under collisions. In
soliton-antisoliton scattering, either both solitary waves may be
annihilated or, alternatively, they may form a bound (breather)
state. Elementary particles share all of these properties. Thus, if
an appropriate system of non-linear field equations admits solitary
waves (or solitons) as solutions, these non-dispersive waves behave
almost as elementary particles.

Nevertheless, the main feature of these solutions, the confinement
of the energy density, has straightforward implementation in many
areas. In Condensed Matter, these solutions describe interfaces in
magnetic materials \cite{Eschen} and in ferroelectric crystals
\cite{Jona}. They have also been used to understand some bizarre
properties of the poly(oxyethylene) \cite{Zhang}, widely studied
owing to its biotechnical and biomedical applications \cite{Harris}.
Among the impressive features of this polymer, we can mention the
ability to entrap metallic ions in aqueous solution and its
solubility in water to almost any extent. On the other hand, in
Cosmology solitary waves are interpreted as domain walls which
seeded the formation of structures in the early universe
\cite{Vilenkin}. They are also of direct interest in high energy
physics, for instance, describing gravity in warped spacetimes
involving $D$ extra spatial dimensions \cite{Koba}, or the coupling
of scalar matter fields with dilaton gravity \cite{Antunes}, which
seems to be related to the formation and evaporation of black holes
\cite{Banks}. Furthermore, when these defects appear as BPS states
in both extended supersymmetric gauge theories \cite{Olive} and
string/M theory \cite{Duff} they play a crucial r\^ole in the
understanding of dualities between the different regimes of the
system. In this framework, they behave as extended states in ${\cal
N}=1$ SUSY gluodynamics and the Wess-Zumino model \cite{Dvali}. The
structure of the solitary waves that we are going to unveil in this
paper is richer than the usual kink structure found in simpler
models. Thus, our topological defects could describe more subtle
effects in each of all these scenarios.

In general, the above mentioned systems involve a high number of
fields. In order to investigate the presence of solitary waves or
topological defects, the usual procedure is to obtain an effective
scalar field theory, carrying out severe restrictions on the
original theory. In most cases one is compelled to pursue an
effective theory that corresponds to a single scalar field model,
where the existence of topological defects can be checked easily. In
general, however, the effective theory depends on several scalar
fields, and the truncation can involve a important loss of
information about the presence of solitary waves or topological
defects. Therefore, it is desirable to investigate the general
properties of solitary waves in a multi-scalar field theory. This is
an important qualitative step as reported by Rajaraman
\cite{Rajaraman}: \textit{This already brings us to the stage where
no general methods are available for obtaining all localized static
solutions, given the field equations. However, some solutions, but
by no means all, can be obtained for a class of such Lagrangians
using a little trial and error}.

Some work on two-component scalar field theory models equipped with
a Minkowskian space-time has been accomplished, see for instance
References \cite{modeloa,Monto1,Ito1,B3,B5,Mar1,Mar,Aai2}. In some
of these models, two-parametric families of solitary waves or kinks
are identified. Generally, each member of these families is a
composite solitary wave, which involves several lumps, i.e, the
energy density is localized at several points. The parameters
identifying each solution correspond to a translational parameter,
$x_0$, which determines the center of the kink, and an \lq\lq orbit"
parameter $b$, which specifies the separation between the lumps.
Study of Manton's adiabatic motion of solitary waves \cite{Manton}
is very simple in this case because the geometric metric of the
moduli (parameter) space that governs the dynamics does not depend
on the translational parameter $x_0$; thus, it is always possible to
apply a transformation that leads to an Euclidean metric. On the
other hand, there are only a few works that have addressed research
into solitary waves in three-component scalar field theory models,
see \cite{Aai3,Aai4,Aai5,BLW}. In this richer case there are
three-parametric families of kinks, and the adiabatic evolution of
three-body lumps is associated with metrics with curvature. The
low-energy dynamics of, in this case, three-body solitary waves is
therefore much more intricate.

In a slightly different physical context, where the solitary waves
are understood as domain walls (2-branes) in (3+1) dimensions models
with three scalar fields have been discussed in Reference \cite{BB}.
Considering topological defects (0-branes) in more spatial
dimensions (p-branes) allows for complex structures having defects
nested inside defects. It is plausible that the composite solitary
waves to be discussed in this paper will give rise to even richer
structures of nested defects inside defects if considered in more
dimensions. An interesting work is also available on networks of
topological defects in a model with two complex scalar fields,  see
\cite{Sut}. The potential energy density is a polynomial in the
fields of arbitrary order chosen in such a way that the model enjoys
a $U(1)\times{\mathbb Z}_n$ symmetry, networks of domain walls
arising in the surface of $Q$-balls. In other direction, defects
inside defects in models with two scalar fields have been considered
before, see \cite{BB1} and \cite{Morris}. Very recently, networks of
topological defects arising in a model with $9n$ scalar fields in
nine spatial dimensions have been studied in \cite{BB2} and the
evolution of these (or similar) structures has been analyzed in
\cite{Av}.

In this paper, we shall address a three-component scalar field
model that generalizes the two-component model discussed in
Reference \cite{modeloa} to three fields. The organization of the
paper is as follows: in Section 2 we introduce the model and
describe the spontaneous symmetry breaking pattern as well as the
spectrum of dispersive waves. In Section 3 we describe the variety
of solitary (non-dispersive) waves that arises in this system. We
classify these solutions according to the number of lumps they are
made of. There are three basic solitary waves with the energy
density confined in one finite region. The rest of the solutions
are composite and display several lumps --the energy density is
localized at various points--.

\section{Deformation of the $\lambda(\vec{\phi}\cdot\vec{\phi})^3_2$-model }

We shall deal with a three-component scalar field theory model
defined in a (1+1) Minkowskian space-time. The dynamics is governed
by the action functional:
\begin{equation}
S=\int d^2 x \left[\frac{1}{2} \partial_\mu \phi^a \partial^\nu
\phi^a - V(\phi_1,\phi_2,\phi_3) \right] \label{eq:action}
\end{equation}
and the potential term is the following sixth-degree polynomial in
the fields:
\begin{equation}
V(\vec\phi)=
(\phi_1^2+\phi_2^2+\phi_3^2)(\phi_1^2+\phi_2^2+\phi_3^2-1)^2+2(\sigma_2^2
\phi_2^2+\sigma_3^2
\phi_3^2)(\phi_1^2+\phi_2^2+\phi_3^2-1)+\sigma_2^4
\phi_2^2+\sigma_3^4 \phi_3^2 \qquad .\label{eq:potential}
\end{equation}
$\sigma_2$ and $\sigma_3$ are the non-dimensional coupling constants
of the system, chosen, without loss of generality, such that
$\sigma_2<\sigma_3$ . $\phi^a$ , $a=1,2,3$, are dimensionless scalar
fields,
\[
\vec{\phi}(x_0,x_1)=(\phi_1(x_0,x_1),\phi_2(x_0,x_1),\phi_3(x_0,x_1)):\mathbb{R}^{1,1}\rightarrow
\mathbb{R}^3 \qquad .
\]
Our convention for the metric tensor components in Minkowski space
${\mathbb R}^{1,1}$ is: $g_{00}=-g_{11}=1$ and $g_{12}=g_{21}=0$.

This model is a $(\sigma_2,\sigma_3)$ deformation of the same system
with $SO(3)$-invariant potential energy density:
\begin{equation}
V_{\rm S}(\vec\phi)=
(\phi_1^2+\phi_2^2+\phi_3^2)(\phi_1^2+\phi_2^2+\phi_3^2-1)^2
\label{eq:potentials}
\end{equation}
Besides the Poincare transformations acting on ${\mathbb R}^{1,1}$,
the non-deformed model -$\sigma_2=\sigma_2=0$- is invariant under
$O(3)$ rotations in internal space ${\mathbb R}^3$. The specific
form of $V_{\rm S}$, however, shows spontaneous symmetry breaking of
the internal symmetry to a $SO(2)$ subgroup so that two Goldstone
bosons are unavoidable if the asymmetric vacuum -any point at a
distance 1 from the origin in ${\mathbb R}^3$- is chosen to perform
the quantization of the system; the set of zeroes of $V_{\rm S}$ is
the the union of a discrete point, the origin, and a continuous
manifold, the 2-sphere of radius 1 in ${\mathbb R}^3$. Coleman
proved in \cite{Co} that there are no Goldstone bosons on the line
in a sensible physical system. The infrared asymptotic behavior of
the quantum system would require modification of the potential
$V_{\rm S}$ in such a way that the zeroes of the potential become a
discrete set and massless particles are forbidden. Our deformation
complies with this requirement and could be understood as being
generated by infrared quantum fluctuations. The added terms in the
potential (\ref{eq:potential}) explicitly break the $O(3)$ symmetry
to the discrete subgroup ${\mathbb G}={\mathbb Z}_2 \times {\mathbb
Z}_2 \times {\mathbb Z}_2$, generated by the reflections $\phi_1
\rightarrow -\phi_1$, $\phi_2 \rightarrow -\phi_2$ and $\phi_3
\rightarrow -\phi_3$ in internal space. We also stress that the
present system is a generalization of the two-component scalar field
theory model discussed in Reference \cite{modeloa}, where the
dimension of the internal space, the number of scalar fields, is
increased by one unit.

The field equations form the system of coupled second-order partial
differential equations:
\begin{eqnarray}
\frac{\partial^2 \phi_1}{\partial t^2}-\frac{\partial^2
\phi_1}{\partial x^2}&=&2\phi_1\left[3
(\phi_1^2+\phi_2^2+\phi_3^2)^2 +1-4 \phi_1^2
-2(2-\sigma_2^2)\phi_2^2
-2 (2-\sigma_3^2)\phi_3^2  \right] \nonumber\\
\frac{\partial^2 \phi_2}{\partial t^2}-\frac{\partial^2
\phi_2}{\partial x^2}&=&2\phi_2\left[3
(\phi_1^2+\phi_2^2+\phi_3^2)^2 -2 (2-\sigma_2^2) \phi_1^2 -4
\bar\sigma_2^2 \phi_2^2 -2 (\bar\sigma_2^2+\bar\sigma_3^2)
\phi_3^2 +\bar\sigma_2^4\right] \label{eq:partial}\\
\frac{\partial^2 \phi_3}{\partial t^2}-\frac{\partial^2
\phi_3}{\partial x^2}&=&2\phi_3\left[3
(\phi_1^2+\phi_2^2+\phi_3^2)^2 -2 (2-\sigma_3^2) \phi_1^2-2
(\bar\sigma_2^2+\bar\sigma_3^2) \phi_2^2 -4 \bar\sigma_3^2 \phi_3^2
+\bar\sigma_3^4\right] \nonumber \qquad .
\end{eqnarray}
For the sake of simplicity, henceforth we use the notation: $x_0=t$,
$x_1=x$, $\bar{\sigma}_2=\sqrt{1-\sigma_2^2}$,
$\bar{\sigma}_3=\sqrt{1-\sigma_3^2}$,
$\sigma^2_{32}=\sigma_3^2-\sigma_2^2$.

\subsection{Structure of the configuration space}

A \lq\lq point" in the configuration space of the system is a
configuration of the field of finite energy; i.e., a picture of the
field at a fixed time such that the energy $E$, the integral over
the line of the energy density,
\begin{equation}
 E[\vec{\phi}]=\int_{-\infty}^\infty dx {\cal E}[\vec\phi]\qquad , \qquad {\cal
E}[\vec{\phi}]=\frac{1}{2} \left( \frac{d\phi_1}{d x} \right)^2
+\frac{1}{2} \left( \frac{d\phi_2}{d x} \right)^2+\frac{1}{2} \left(
\frac{d\phi_3}{d x} \right)^2+V(\phi_1,\phi_2,\phi_3) \qquad ,
\label{eq:density}
\end{equation}
is finite. Thus, the configuration space is the set of continuous
maps from ${\mathbb R}$ to ${\mathbb R}^3$ of finite energy:
\[
{\cal C}=\left\{\vec{\phi}(x)\in {\rm Maps}({\mathbb R}, {\mathbb
R}^3)/E<\infty\right\} \qquad .
\]
In order to belong to ${\cal C}$, each configuration complies with
the asymptotic conditions
\begin{equation}
\lim_{x\rightarrow \pm \infty} \vec{\phi} \in {\cal M} \hspace{2cm}
\lim_{x\rightarrow \pm \infty} \frac{d\vec{\phi}}{dx} =0
\label{eq:asymtotic}
\end{equation}
where ${\cal M}$ is the set of zeroes (minima) of the potential term
$V(\vec{\phi})$. In our model, ${\cal M}$ consists of seven
elements, see Figure 1:
\begin{equation}
{\cal M}=\left\{ \bar{\phi}^{O}\equiv (0,0,0); \hspace{0.2cm}
\bar{\phi}^{D^\pm}\equiv (\pm 1,0,0); \hspace{0.2cm}
\bar{\phi}^{B^\pm}\equiv ( 0,0,\pm \bar\sigma_3); \hspace{0.2cm}
\bar{\phi}^{C^\pm}\equiv (0,\pm \bar\sigma_2,0) \right\}
\label{eq:vac}
\end{equation}
Note that all the zeroes except the origin lie in the ellipsoid
${\mathbb E}\equiv \phi_1^2 + \frac{\phi_2^2}{\bar\sigma_2^2}+
\frac{\phi_3^2}{\bar\sigma_3^2}=1$. The intersections of the
ellipsoid ${\mathbb E}$ and the $\phi_1=0$, $\phi_2=0$,
$\phi_3=0$-planes are respectively the ellipses $e_1 \equiv
\frac{\phi_2^2}{\bar\sigma_2^2}+ \frac{\phi_3^2}{\bar\sigma_3^2}=1$,
$e_2\equiv \phi_1^2 + \frac{\phi_3^2}{\bar\sigma_3^2}=1$ and
$e_3\equiv \phi_1^2 + \frac{\phi_2^2}{\bar\sigma_2^2}=1$ of
eccentricities $\varepsilon (e_1)=\frac{\sigma_{32}^2}{\sigma_2^2}$,
$\varepsilon(e_2)=\sigma_3^2$ and $\varepsilon(e_3)=\sigma_2^2$. The
minima $B_\pm$  are on the $\phi_3$-axis at the intersection between
$e_1$ and $e_2$; $C_\pm$ lies on the $\phi_2$-axis at $e_1 \cap e_3
$, and $D_\pm$ on the $\phi_1$-axis at $e_2 \cap e_3$.

\begin{figure}[htb]
\centerline{\includegraphics[height=3.cm]{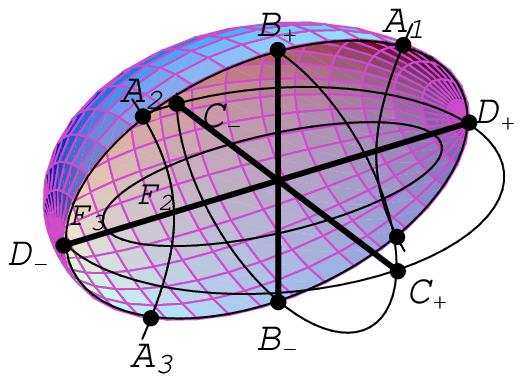}\hspace{2cm}
\includegraphics[height=3.cm]{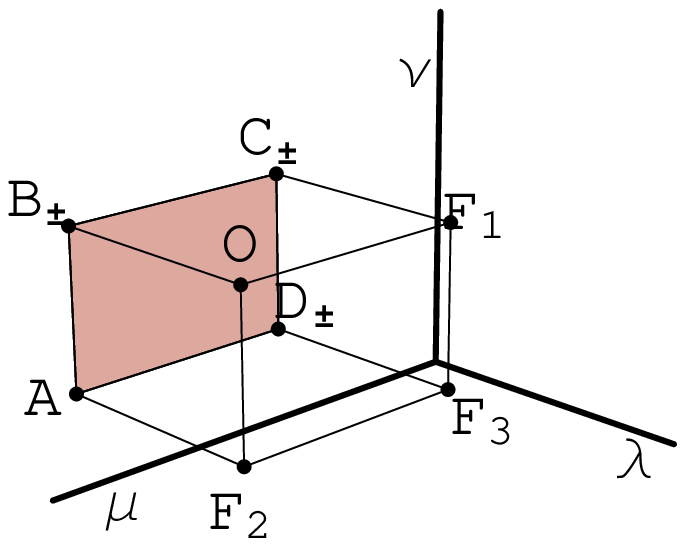}}
\caption{\small \textit{Points belonging to ${\cal M}$, umbilical
points, focal points, and characteristic curves of the model drawn
on and inside the ellipsoid ${\mathbb E}$: (left) Cartesian
coordinates. (right) Elliptic coordinates.}}
\end{figure}

The configuration space ${\cal C}$ is the union of 49 topologically
disconnected sectors:
\[
{\cal C}=\cup_{I,J} {\cal C}^{IJ}
\]
where $I,J=\{O,D,C,B\}$ and ${\cal C}^{IJ}$ stands for the set of
configurations that connect the points $\bar{\phi}^{I}$ with
$\bar{\phi}^{J}$. Because temporal evolution is a homotopy
transformation, the asymptotic conditions (\ref{eq:asymtotic}) do
not change with $t$ and the sectors are completely disconnected.
Physically, this means that a configuration in a sector cannot
evolve into configurations in other different sectors; it would cost
infinite energy.

\subsection{Plane waves and spontaneous symmetry breakdown}

The zeroes of the potential energy density (\ref{eq:vac}) are also
homogeneous static solutions of the field equations
(\ref{eq:partial}) with zero energy; in fact, they are absolute
minima of the energy in the topological sectors ${\cal C}^{II}$ . A
spontaneous symmetry breaking arises because of the degeneracy of
the set ${\cal M}$. Whereas the point $\bar{\phi}^O$ located in the
origin preserves the discrete symmetry ${\mathbb Z}_2 \times
{\mathbb Z}_2 \times {\mathbb Z}_2$ of the Lagrangian, the choice of
$\bar{\phi}^{D^\pm}$ breaks the symmetry generated by the reflection
$\phi_1 \rightarrow -\phi_1$, which connects the two points,
$\bar{\phi}^{D^+}$ and $\bar{\phi}^{D^-}$, to the little group
$\{e\} \times {\mathbb Z}_2 \times {\mathbb Z}_2$. Similar
considerations work for the pair of points $\bar{\phi}^{C^\pm}$ and
$\bar{\phi}^{B^\pm}$ with respect to the transformations $\phi_2
\rightarrow -\phi_2$ and $\phi_3 \rightarrow -\phi_3$. The moduli
space of vacua comprises four ${\mathbb G}$-orbits
\[
\bar{{\cal M}}=\frac{{\cal M}}{\mathbb G}=\{\, \bar{\phi}^{O}\,; \,
\bar{\phi}^{D}\, ; \,\bar{\phi}^{B}\, ; \, \bar{\phi}^{C}\, \}
\]
where we write
$\bar{\phi}^{D}=\{\bar{\phi}^{D^+},\bar{\phi}^{D^-}\}$,
$\bar{\phi}^{C}=\{\bar{\phi}^{C^+},\bar{\phi}^{C^-}\}$ and
$\bar{\phi}^{B}=\{\bar{\phi}^{B^+},\bar{\phi}^{B^-}\}$.

Small fluctuations around the homogenous static solutions
$\bar\phi$, $\psi(t,x)=\bar\phi + \delta \psi (t,x)$ are still
solutions of (\ref{eq:partial}) if the linear PDE system
\begin{equation}
\sum_{b=1}^3 \left( \Box \delta_{ab} + M_{ab}^2(\bar\phi) \right)
\delta \psi_b(t,x)=0_a \, , \hspace{1cm} M_{ab}^2=\frac{\partial^2
V}{\partial \phi_a \partial \phi_b} (\bar{\phi}) \label{eq:spectral}
\end{equation}
is satisfied. The solution of (\ref{eq:spectral}) via the separation
of variables $\delta \psi_a(t,x) ={\rm Re}\left( e^{i\omega
t}\right) f_a^\omega (x)$ leads to the spectral problem for the
second-order fluctuation --or Hessian-- operator:
\begin{equation}
\sum_{b=1}^3 {\cal H}_{ab}(\bar\phi) f_b^\omega (x)=\sum_{b=1}^3
\left(-\frac{d^2}{dx^2} \delta_{ab} + M^2_{ab}(\bar\phi)\right)
f_b^\omega (x)=\omega^2 f_a^\omega (x) \label{eq:spectral2}
\end{equation}
${\cal H}(\bar{\phi})$ is a diagonal matrix differential operator
for the four points of $\bar{\cal M}$ with a positive definite
spectrum; each constant solution belonging to ${\cal M}$ is stable.
Thus, there are four types of dispersive wave-packet solutions, each
one with three branches
\[
f_1^\omega(x)={\rm sin}\,kx \, , \, f_2^\omega=f_3^\omega=0 \quad ;
\quad f_1^\omega=0 \, , \, f_2^\omega(x)={\rm sin}\, qx \, , \,
f_3^\omega=0 \quad ; \quad f_1^\omega=f_2^\omega=0 \, , \,
f_3^\omega(x)={\rm sin}\, px \quad ,
\]
living in one of these sectors. The building blocks are these plane
waves and the dispersion laws of the corresponding wave packets are
respectively determined by the diagonal matrices:
\[
M^2(\bar\phi^{O})=\left( \begin{array}{ccc} 2 & 0 & 0 \\ 0 &
2\bar\sigma_2^2 & 0
\\ 0 & 0 & 2\bar\sigma_3^2
\end{array} \right)\hspace{1.5cm} M^2(\bar\phi^{D})=
\left( \begin{array}{ccc} 8 & 0 & 0 \\ 0 & 2\sigma_2^4 & 0
\\ 0 & 0 & 2 \sigma_3^4
\end{array} \right)
\]
\[
M^2(\bar\phi^{C})= \left( \begin{array}{ccc} 2 \sigma_3^4 & 0 & 0
\\ 0 & 8 \bar\sigma_2^4 & 0
\\ 0 & 0 & 2 \sigma_{32}^4
\end{array} \right) \hspace{1.5cm} M^2(\bar\phi^{B})= \left( \begin{array}{ccc} 2 \sigma_2^4 & 0 &
0 \\ 0 & 2 \sigma_{32}^4 & 0
\\ 0 & 0 & 8 \bar\sigma_3^4
\end{array} \right)
\]
The three branches for each type are:

\[
\begin{array}{|cc|cc|c|cc|} \hline &&&&&&\\[-0.3cm]  \bar{\phi}^D & & \bar{\phi}^C & & \bar{\phi}^B & &
\bar{\phi}^O \\ \hline &&&&&&\\[-0.3cm] \omega^2(k)=k^2+8 & &
\omega^2(k)=k^2+2\sigma_3^4 & &
\omega^2(k)=k^2+2\sigma_2^4 & & \omega^2(k)=k^2+2 \\
\omega^2(q)=q^2+2\sigma_2^4 & & \omega^2(q)=q^2+8\bar{\sigma}_2^4
& & \omega^2(q)=q^2+2\sigma_{32}^4 & & \omega^2(q)=q^2+2\bar{\sigma}_2^2 \\
\omega^2(p)=p^2+2\sigma_3^4 & & \omega^2(p)=p^2+2\sigma_{32}^4 & &
\omega^2(p)=p^2+8\bar{\sigma}_3^4 & &
\omega^2(p)=p^2+2\bar{\sigma}_3^2
\\
\hline
\end{array}
\]

\subsection{Solitary waves from integrable dynamical systems}

According to Rajaraman \cite{Rajaraman} \lq\lq A solitary wave is a
localized non-singular solution of any non-linear field equation
whose energy density, as well as being localized, has space-time
dependence of the form: $\varepsilon(t,x)=\varepsilon(x-\rm{v} t)$,
where v is some velocity vector". Because we are dealing with a
Lorentz invariant system, solutions with a temporal dependence like
this are obtained from static solutions by means of a Lorentz
velocity transformation:\hspace{0.6cm} $L({\rm v})\phi(x)=
\phi^L(t,x)=\phi\left( \frac{x-{\rm v} t}{\sqrt{1-{\rm v}^2}}
\right)$.

Thus, the search for solitary waves is tantamount to looking for
localized solutions in the ${\cal C}^{IJ}, I\neq J$ sector of the
ODE system :
\begin{equation}
\frac{d^2 \phi_a}{dx^2}= \frac{\partial V}{\partial \phi_a}
\hspace{1cm} a=1,2,3 \label{eq:ordinary} \qquad .
\end{equation}
Understanding the scalar field $\vec\phi$ as the \textit{particle
position vector}; $x$ as the \textit{particle time} and $U=-V$ as
the \textit{particle potential}, this is no more Newton equations
describing the motion of a particle in three dimensions in the field
of the potential $U$. Moreover, the energy functional and the energy
density (\ref{eq:density}) of the field theory turns respectively
into the action and the Lagrangian of the mechanical system. The
Legendre transformation provides the Hamiltonian of the analogous
mechanical system:
\begin{equation}
H=\sum_{a=1}^3 p_a \frac{d\phi_a}{dx}-{\cal
E}[\vec\phi,\frac{d\vec\phi}{dx}]={1\over 2}\sum_{a=1}^3 p_a(x)\cdot
p_a(x)-V(\vec{\phi}(x)) \qquad \quad , \qquad \quad
p_a=\frac{\partial {\cal E}}{\partial \phi^a /
\partial x} \, \, \, \label{eq:ham} \quad .
\end{equation}
The Hamiltonian (\ref{eq:ham}) corresponds to a St\"ackel separable
system. To prove this, we introduce three-dimensional Jacobi
elliptic coordinates $(\lambda,\mu,\nu)$ in the internal space
$(\phi_1,\phi_2,\phi_3)$, see \cite{Aai3}.  In terms of elliptic
coordinates, the fields are:
\begin{equation}
\phi_1^2=\frac{(1-\lambda)(1-\mu)(1-
\nu)}{(1-\bar\sigma_3^2)(1-\bar\sigma_2^2)} ;\hspace{0.5cm}
\phi_2^2=
\frac{(\bar\sigma_2^2-\lambda)(\bar\sigma_2^2-\mu)(\bar\sigma_2^2-
\nu)}{(\bar\sigma_2^2-\bar\sigma_3^2)(\bar\sigma_2^2-1)};\hspace{0.5cm}
\phi_3^2=
\frac{(\bar\sigma_3^2-\lambda)(\bar\sigma_3^2-\mu)(\bar\sigma_3^2-
\nu)}{(\bar\sigma_3^2-\bar\sigma_2^2)(\bar\sigma_3^2-1)} \qquad .
\label{eq:elipticas}
\end{equation}
The new variables take values in the infinite cuboid
\[
-\infty \leq \lambda \leq \bar\sigma_3^2 \leq \mu \leq
\bar\sigma_2^2 \leq \nu<1
\]
which is mapped by the coordinate transformation
(\ref{eq:elipticas}) to a Cartesian octant in ${\mathbb R}^3$.
Therefore eight points in ${\mathbb R}^3$ become only one in the
cuboid and mapping back solutions in elliptic coordinates to the
original Cartesian ones requires a lot of care to reassemble the
solution in the total internal space from one octant by imposing
continuity of the trajectories and their derivatives. Apparent
rebounds on the faces of $P(0)$, see Figure 1, are really continuous
in the closed domain in ${\mathbb R}^3$ bounded by the ellipsoid
${\mathbb E}$.

In Figure 1 we have plotted the finite elliptic cuboid, denoted by
$P(0)$, where (an octant of) the interior of the ellipsoid ${\mathbb
E}$ is mapped. The ellipsoid itself is mapped to the $\lambda=0$
face of $P(0)$, ${\mathbb E} \equiv \lambda=0$. In elliptic
coordinates, the following information is interesting:
\begin{itemize}
\item Homogeneous solutions
\begin{eqnarray*}
\bar{\phi}^O \equiv (\lambda=\bar{\sigma}_3^2 , \mu=\bar{\sigma}_2^2
, \nu=1) \qquad &,& \bar{\phi}^D \equiv
(\lambda=0 , \mu=\bar{\sigma}_2^2 , \nu=\bar{\sigma}_3^2) \\
\bar{\phi}^B \equiv (\lambda=0 , \mu=\bar{\sigma}_3^2 , \nu=1)
\qquad &,& \bar{\phi}^C \equiv (\lambda=0 , \mu=\bar{\sigma}_3^2 ,
\nu=1)
\end{eqnarray*}

\item Ellipses where the homogeneous solutions live:
\[
e_1 \equiv \lambda=0 \, , \, \nu=1 \hspace{1cm} ; \quad e_2\equiv
\left\{\begin{array}{ccc}\lambda=0 & , & \mu=\bar{\sigma}_2^2 \\
\lambda=0 & , & \nu=\bar{\sigma}_2^2\end{array}\right. \hspace{1cm}
; \quad  e_3\equiv \lambda=0 \, , \, \mu=\bar{\sigma}_3^2
\]

\item Other special points: $F_1$, $F_2$, $F_3$ (foci of $e_1$, $e_2$ and $e_3$
respectively) and the umbilicus points $A$ of the ellipsoid
${\mathbb E}$ are settled on the vertices of the parallelepiped
$P(0)$ in elliptic space:
\begin{eqnarray*}
F_1 \equiv (\lambda=\bar{\sigma}_3^2 , \mu=\bar{\sigma}_3^2 , \nu=1)
\qquad &,& \qquad F_2 \equiv
(\lambda=\bar{\sigma}_3^2 , \mu=\bar{\sigma}_2^2 , \nu=\bar{\sigma}_2^2) \\
F_3 \equiv (\lambda=\bar{\sigma}_3^2 , \mu=\bar{\sigma}_3^2 ,
\nu=\bar{\sigma}_2^2) \qquad &,& \qquad A \equiv (\lambda=0 ,
\mu=\bar{\sigma}_2^2 , \nu=\bar{\sigma}_2^2)
\end{eqnarray*}

\item Special curves such as the ellipse $e_4 \equiv
\frac{\phi_1^2}{\sigma_3^2}+\frac{\phi_2^2}{\sigma_{32}^2}=1$
corresponding to the edge $F_1F_3$ in $P(0)$ and the hyperbola $h
\equiv\frac{\phi_1^2}{\sigma_2^2}-\frac{\phi_3^2}{\sigma_{32}^2}=1$,
the edge $F_2A$.
\[
e_4 \equiv \hspace{0.5cm}\lambda=\bar{\sigma}_3^2 \, , \,
\nu=\bar{\sigma}_2^2 \qquad , \qquad h \equiv
\hspace{0.5cm}\mu=\bar{\sigma}_2^2 \, , \, \nu=\bar{\sigma}_2^2
\qquad .
\]
We shall see that these curves are focal lines of finite action
trajectories in the analogous mechanical system.
\end{itemize}

In elliptic coordinates, the Hamiltonian (\ref{eq:ham}) of the
analogous mechanical system becomes:
\begin{equation}
H=\frac{H_\lambda}{(\lambda-\mu)(\lambda-\nu)}+
\frac{H_\mu}{(\mu-\lambda)(\mu-\nu)}+
\frac{H_\nu}{(\nu-\lambda)(\nu-\mu)} \qquad . \label{eq:hamiltonian}
\end{equation}
Here,
\[
H_\lambda=2 P_3(\lambda)p_\lambda^2 + P_4(\lambda);\hspace{1cm}
H_\mu=2 P_3(\mu)p_\mu^2 + P_4(\mu);\hspace{1cm}H_\nu=2
P_3(\nu)p_\nu^2 + P_4(\nu)
\]
\[
P_3(x)=(1-x)(\bar\sigma_2^2-x)(\bar\sigma_3^2-x) \hspace{1cm}
P_4(x)=x^2(x-1)(x-\bar\sigma_2^2)(x-\bar\sigma_3^2)
\]
Expression (\ref{eq:hamiltonian}) is of the St\"ackel form. Thus,
the analogous mechanical system is Hamilton-Jacobi separable. We
will take advantage of this fact in the following Sections in order
to identify the variety of solitary waves in the model as the finite
action trajectories of the analogous mechanical system.

\section{The variety of solitary waves}

In this Section we describe the variety of solitary waves in
successive stages. The analogous mechanical system encompasses a
vast manifold of finite action trajectories in one-to-one
correspondence with the solitary waves of the field theoretical
model. One-body solitary waves, made of only one lump, can be
found by means of Rajaraman's trial orbit method, without the need
to use elliptic coordinates. On plugging some specific trial
curves into the equations (\ref{eq:ordinary}), simple solitary
waves that we will refer to as basic lumps arise. The trial orbits
correspond to some edges of the parallelepiped $P(0)$ shown in
Figure 1(b). Later, solutions which comprise several of the basic
lumps will be found by restricting the movement to the faces of
the parallelepiped $P(0)$ to find solitary waves akin to those
discovered in Reference \cite{modeloa} . As a last step, we shall
solve the equations (\ref{eq:ordinary}) in the general case by
applying the Hamilton-Jacobi theory with no restrictions.

The system of ODE (\ref{eq:ordinary}) is invariant under
translations $x\rightarrow x-x_0$ and reflections $x\rightarrow
-x$ in the spatial parameter $x$. The first symmetry allows us to
localize the center of the solitary wave or kink at an arbitrary
point $x_0$, whereas the second symmetry connects a kink with its
antikink. Henceforth, we shorthand these transformations on $x$ as
$\bar{x}$, i.e, $\bar{x}=(-1)^\delta (x-x_0)$ with
$x_0\in\mathbb{R}$ and $\delta=0,1$.

\subsection{Solitary waves: basic types}

\textit{1. $K_1^{OB}$:} The solutions of (\ref{eq:ordinary}) in
the sector $C^{OB}$, located in the $\phi_3$-axis in internal
space, are four-fold, see Figure 2(c):
\begin{equation}
\bar\phi_1(x)= 0 \, ; \hspace{1cm} \bar\phi_2(x)=0 \, ;
\hspace{1cm} \bar\phi_3(x)=\frac{(-1)^\alpha
\bar\sigma_3}{\sqrt{1+e^{2\sqrt{2}\bar\sigma_3^2 \bar x}}}
\label{eq:kob}
\end{equation}
Here, $\alpha=0,1$ distinguishes between solutions with $\phi_3>0$
and $\phi_3<0$. We refer to these solutions as $K_1^{OB}$. Some
remarks about this notation: the superscripts specify the elements
of $\bar{\cal M}$ that are connected by the solution and/or the
topological sector ${\cal C}^{OB}$, where the solitary wave lives.
The subscript shows the number of basic particles, lumps or bodies
of which a particular solitary wave is made. In Figure 2(a) a
$\alpha=\delta=0$ $K_1^{OB}$ solitary wave of kink shape is
depicted; this $K_1^{OB^+}$ kink connects the points $B_+$ and $O$
in $\bar{\cal M}$ and lives in the ${\cal C}^{OB^+}$ topological
sector of the configuration space. In Figure 2(b) its energy
density is plotted as a function of $x$; because it is localized
at a single point we understand these one-body solutions as basic
(non-composite) particles. A Lorentz transformation applied to
(\ref{eq:kob}) will show this lump of energy to be a solitary
(traveling) wave. Integration on the real line provides us with
the total energy, which for these solutions is
$E(K_1^{OB})=\frac{\bar\sigma_3^4}{2\sqrt{2}}$ and this depends
only on the coupling constant $\sigma_3$.

\begin{figure}[hbt]
\centerline{\includegraphics[height=3cm]{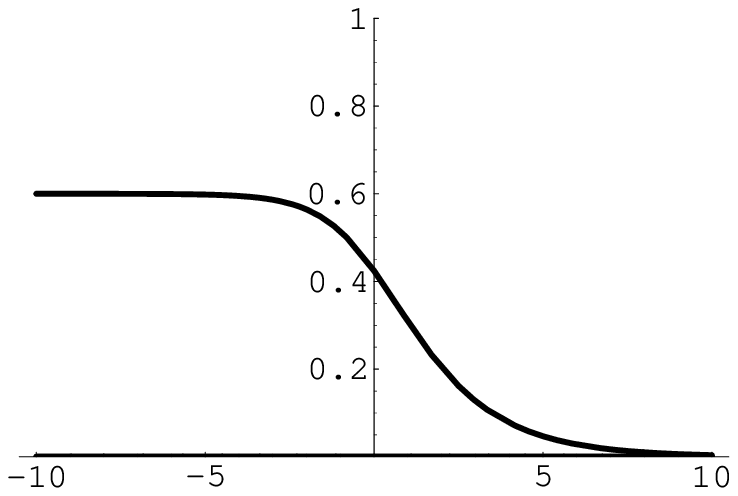}
\hspace{0.6cm}
\includegraphics[height=3cm]{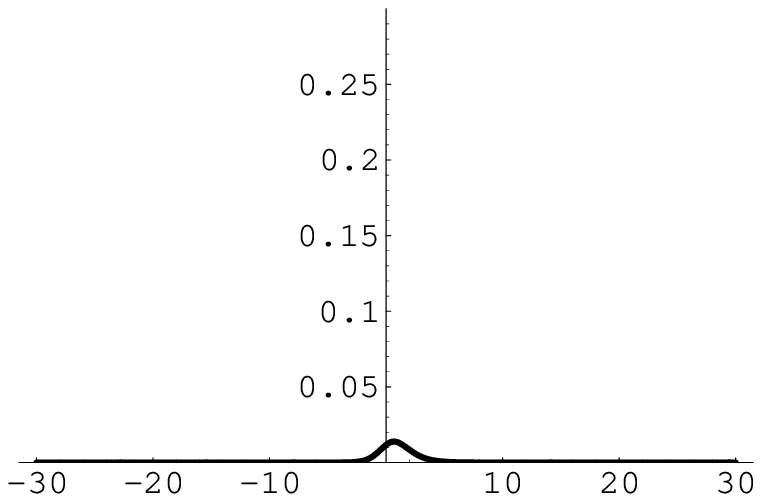} \hspace{0.6cm}
 \includegraphics[height=3cm]{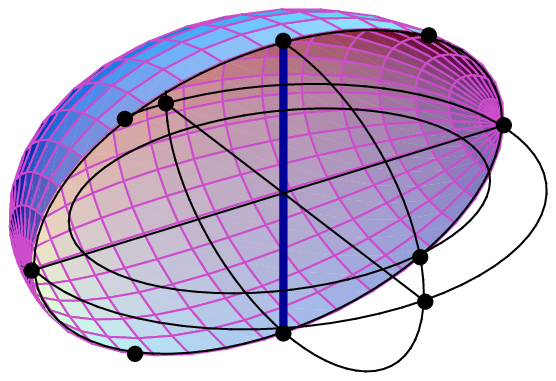}}
\caption{\small \textit{Kink form factor (a), energy density (b)
and orbits (c) for $K_1^{OB}$ kinks. }}
\end{figure}

We shall adopt the policy of showing all the energy density plots
with the same vertical range in order to compare the size of the
lumps associated with different solitary waves.

\textit{2. $K_1^{CD}$:} Plugging the trial elliptic orbit
$e_3\equiv \phi_1^2 +\frac{\phi_2^2}{\bar\sigma_2^2}=1$ into the
equations (\ref{eq:ordinary}) we find the eight solutions
\begin{equation}
\bar\phi_1(x)= \frac{(-1)^\alpha}{\sqrt{1+e^{2\sqrt{2}\sigma_2^2
\bar{x}}}} \, ; \hspace{1cm} \bar\phi_2(x)= \frac{(-1)^\beta
\bar\sigma_2}{\sqrt{1+e^{-2\sqrt{2}\bar\sigma_2^2
 \bar x}}}\, ; \hspace{1cm} \bar\phi_3(x)=0
\label{eq:kcd}
\end{equation}
with $\alpha,\beta,\delta=0,1$.  These trajectories are embedded
in the $\phi_3=0$ plane and live in one of the eight ${\cal
C}^{CD}$ topological sectors, see Figure 3(c). Note that both
non-null components, $\phi_1$ and $\phi_2$, have kink shape.

The value of $\delta$ determines whether the solution is a kink or
antikink. The values of $\alpha$ and $\beta$, however,
characterize the quarter-ellipse in the intersection of the
ellipsoid ${\mathbb E}$ with the $\phi_3=0$ plane, where a
particular solution is located; therefore, $\alpha$ and $\beta$
also choose the points that are asymptotically connected by these
kinks. In Figure 3(a) a particular kink $K_1^{CD}$ with
$\alpha=\beta=\delta=0$ is shown that connects the $C_+$ and $D_+$
points. The energy of these kinks is $E(K_1^{CD})=\frac{\sigma_2^2
(2-\sigma_2^2)}{2\sqrt{2}}$, a function of the coupling constant
$\sigma_2$ only. Again, the energy density is localized at one
point, and they are basic particles in our model, see Figure 3(b).

\begin{figure}[hbt]
\centerline{\includegraphics[width=4.5cm]{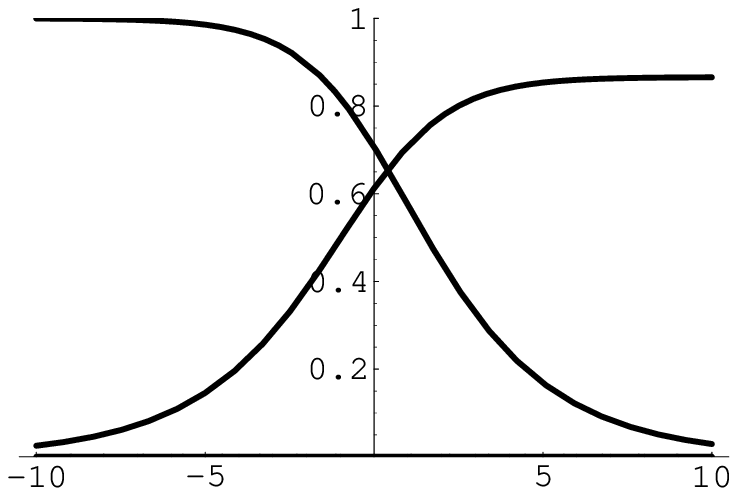}
\hspace{0.6cm}
\includegraphics[width=4.5cm]{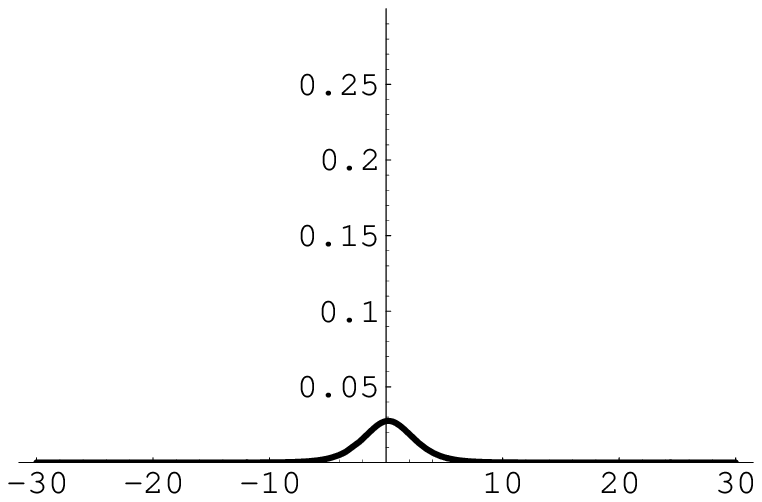} \hspace{0.6cm}
 \includegraphics[width=4.5cm]{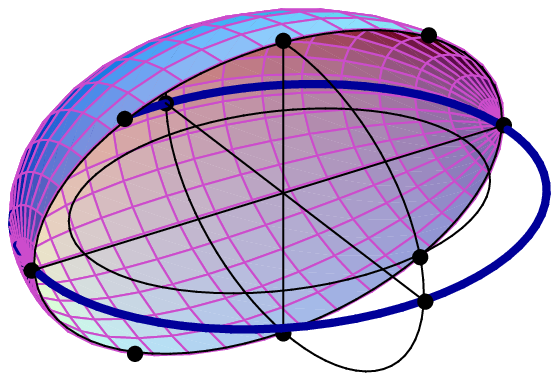}}
\caption{\small \textit{Kink form factor (a), energy density (b)
and orbits (c) for $K_1^{CD}$ kinks. }}
\end{figure}

\textit{3. $K_1^{BC}$:} We shall also try the elliptic orbit
$e_1\equiv
\frac{\phi_2^2}{\bar\sigma_2^2}+\frac{\phi_3^2}{\bar\sigma_3^2}=1$
confined to the plane $\phi_1=0$. Substituting this condition in
(\ref{eq:ordinary}), the following eight solutions, connecting $B$
and $C$, are obtained:
\begin{equation}
\bar\phi_1(x)= 0 \, ; \hspace{1cm} \bar\phi_2(x)=\frac{(-1)^\alpha
\bar\sigma_2 }{\sqrt{1+e^{-2\sqrt{2}\sigma_{32}^2  \bar x}}} \, ;
\hspace{1cm} \bar\phi_3(x)=\frac{(-1)^\beta
\bar\sigma_3}{\sqrt{1+e^{2\sqrt{2} \sigma_{32}^2  \bar x}}}
\label{eq:kbc}
\end{equation}
with $\alpha,\beta,\delta=0,1$. Again the non-null components,
$\phi_2$ and $\phi_3$ in this case, are kink shaped. In Figure
4(a) a $K_1^{BC}$ kink is represented for $\alpha=\beta=\delta=0$.
The analogous trajectory leaves the instability point $B_+$ of
$U=-V$ \lq\lq at" $x=-\infty$ and arrives at another instability
point $C_+$ \lq\lq at" $x=\infty$. These solutions are made of a
single lump and carry energy equal to
$E(K_1^{BC})=\frac{\bar\sigma_2^4-\bar\sigma_3^4}{2\sqrt{2}}$, a
function of both coupling constants.
\begin{figure}[hbt]
\centerline{\includegraphics[height=3cm]{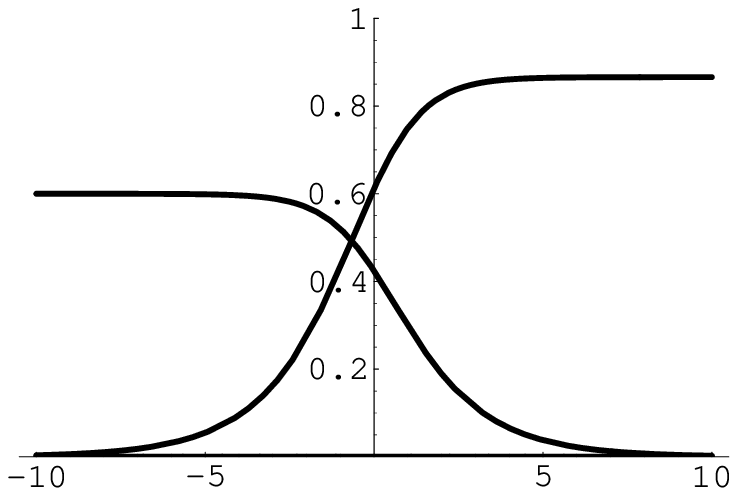}
\hspace{0.6cm}
\includegraphics[height=3cm]{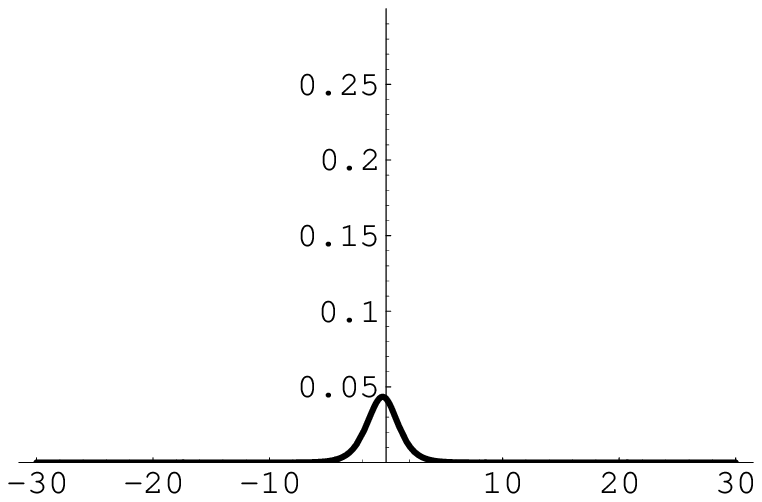} \hspace{0.6cm}
 \includegraphics[height=3cm]{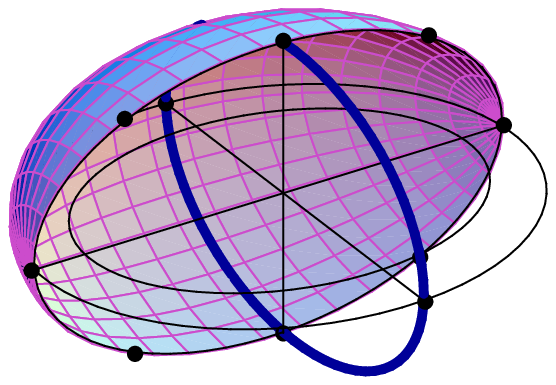}}
\caption{\small \textit{Kink form factor (a), energy density (b)
and orbits (c) for $K_1^{BC}$ kinks. }}
\end{figure}

In sum, there exist three kinds of one-body solitary waves or
basic extended particles, denoted as $K_1^{CD}$, $K_1^{OB}$ and
$K_1^{BC}$. The energy of $K_1^{CD}$ increases as the eccentricity
of the ellipse $e_3$ grows, whereas the energy of $K_1^{OB}$
decreases when the eccentricity of $e_2$ increases. The energy for
the $K_1^{BC}$ lumps depends on the difference of the eccentricity
of the ellipses $e_3$ and $e_2$ and becomes zero when these
magnitudes are equal. Which basic particle is more energetic
depends on the values of the coupling constants $\sigma_2$ and
$\sigma_3$. The figures in this paper have been drawn for the
values $\sigma_2=0.5$ and $\sigma_3=0.8$.

\subsection{Two-body solitary waves}

\textit{1. $K_2^{OC}(b)$:} We shall now tackle the search for
solutions in the $\phi_1=0$ plane, $\nu=1$ in elliptic
coordinates, see Figure 5(c). The Hamiltonian
(\ref{eq:hamiltonian}) restricted to this plane reads
\[
H=\frac{2 P_3(\lambda)p_\lambda^2
+P_4(\lambda)}{(\lambda-\mu)(\lambda-1)}+ \frac{2 P_3(\mu)p_\mu^2
+P_4(\mu)}{(\mu-\lambda)(\mu-1)} \qquad .
\]
The standard Hamilton-Jacobi procedure prescribes how to solve the
HJ equation
\[
\frac{\partial {\cal S}}{\partial x}+H\left( \frac{\partial {\cal
S}}{\partial \lambda},\frac{\partial {\cal S}}{\partial
\mu},\lambda,\mu \right)=0
\]
assuming the separable form of the Hamilton's principal function
${\cal S}={\cal S}_x+{\cal S}_\lambda(\lambda)+{\cal
S}_\mu(\mu)$. The PDE HJ equation becomes a ODE system and the
complete integral can be found by quadratures
\[
{\cal S}_x=-E x;\hspace{0.5cm} {\cal S}_\lambda={\rm s}_\lambda
\int d\lambda
\sqrt{\frac{F-\frac{1}{2}E\lambda}{(\bar\sigma_2^2-\lambda)(\bar\sigma_3^2-\lambda)}
+\frac{\lambda^2}{2}};\hspace{0.5cm} {\cal S}_\mu= {\rm s}_\mu
\int d\mu
\sqrt{\frac{F-\frac{1}{2}E\mu}{(\bar\sigma_2^2-\mu)(\bar\sigma_3^2-\mu)}
+\frac{\mu^2}{2}}
\]
in terms of the integration constants $E$, $F$. ${\rm s}_i ={\rm
Sign}\, p_i$ denote the sign of the momenta.

The trajectories are determined from the orbits selected by the
equation $\gamma_2=\frac{\partial {\cal S}}{\partial F}$ where
$\gamma_2$ is a constant, and the time-schedule fixed by the
equation $\gamma_1=\frac{\partial {\cal S}}{\partial E}$ where
$\gamma_1$ is another arbitrary constant. The asymptotic
conditions (\ref{eq:asymtotic}) picking finite action trajectories
force $E=F=0$. The finite action trajectories of the analogous
mechanical problem are thus determined by the pair of equations
\[
e^{\sqrt{2}\gamma_2}=\left[ \frac{\lambda^{\frac{1}{\bar\sigma_2^2
\bar\sigma_3^2}}(\bar\sigma_2^2-\lambda)^\frac{1}{\bar\sigma_2^2
\sigma_{32}^2}}{(\bar\sigma_3^2-\lambda)^\frac{1}{\bar\sigma_3^2
\sigma_{32}^2 }} \right]^{{\rm s}_\lambda} \left[
\frac{\mu^{\frac{1}{\bar\sigma_2^2
\bar\sigma_3^2}}(\mu-\bar\sigma_2^2)^\frac{1}{\bar\sigma_2^2
\sigma_{32}^2}}{(\bar\sigma_3^2-\mu)^\frac{1}{\bar\sigma_3^2
\sigma_{32}^2 }} \right]^{{\rm s}_\mu} \!\!\! ;\,\, e^{2
\sqrt{2}\sigma_{32}^2 \bar{x}}=
\left[\frac{\bar\sigma_2^2-\lambda}{\bar\sigma_3^2-\lambda}
\right]^{{\rm s}_\lambda}
\left[\frac{\bar\sigma_2^2-\mu}{\mu-\bar\sigma_3^2} \right]^{{\rm
s}_\mu}
\]
which can be reshuffled in the form:
\begin{equation}
e^{-\sqrt{2} \bar\sigma_2^2 (2\bar{x}-\bar\sigma_3^2
\gamma_2)}=\left[ \frac{\lambda}{\bar\sigma_2^2-\lambda}
\right]^{{\rm s}_\lambda} \left[ \frac{\mu}{\bar\sigma_2^2-\mu}
\right]^{{\rm s}_\mu} \!\!; \hspace{0.3cm} e^{-\sqrt{2}
\bar\sigma_3^2 (2\bar{x}-\bar\sigma_2^2 \gamma_2)}=\left[
\frac{\lambda}{\bar\sigma_3^2-\lambda} \right]^{{\rm s}_\lambda}
\left[ \frac{\mu}{\mu-\bar\sigma_3^2} \right]^{{\rm s}_\mu}
\label{eq:nu0}
\end{equation}
The choice ${\rm s}_\lambda={\rm s}_\mu$ in (\ref{eq:nu0}) yields
solutions in the topological sector ${\cal C}^{OC}$. Translating
(\ref{eq:nu0}) back to Cartesian coordinates, we obtain the
following field profiles for the associated one-parametric family
of solitary waves parametrized by $b^2=e^{-\sqrt{2} \sigma_{32}^2
\bar\sigma_3^2 \gamma_2}$:
\begin{equation}
\bar\phi_1= 0; \hspace{0.3cm} \bar\phi_2= \frac{(-1)^\alpha
\bar\sigma_2}{\sqrt{1+ \frac{ \sigma_{32}^2}{\bar\sigma_3^2} e^{-2
\sqrt{2} \bar\sigma_2^2 \bar{x}}+
\frac{\bar\sigma_2^2}{\bar\sigma_3^2} b^2 e^{-2\sqrt{2}
\sigma_{32}^2 \bar x} }} ; \hspace{0.3cm} \bar\phi_3= \frac{b
\bar\sigma_3}{\sqrt{b^2+ \frac{ \sigma_{32}^2}{\bar\sigma_2^2}
e^{-2 \sqrt{2} \bar\sigma_3^2 \bar{x}}+
\frac{\bar\sigma_3^2}{\bar\sigma_2^2} e^{2\sqrt{2} \sigma_{32}^2
\bar x} }} \qquad . \label{eq:k2bc}
\end{equation}
We refer to these solutions as $K_2^{OC}(b)$ because they are made
of two basic lumps. $\alpha=0$ describes solutions joining $O$ and
$C^+$ whereas $\alpha=1$ gives solitary waves which connect the
points $O$ and $C^-$. In Figure 5(a) a $K_2^{OC}$ kink with
$\alpha=\gamma=0$ is depicted. The analogous trajectory departs
from $O$ and arrives at $C_+$ as $x$ goes from $-\infty$ to
$\infty$.

\begin{figure}[hbt]
\centerline{\includegraphics[height=3cm]{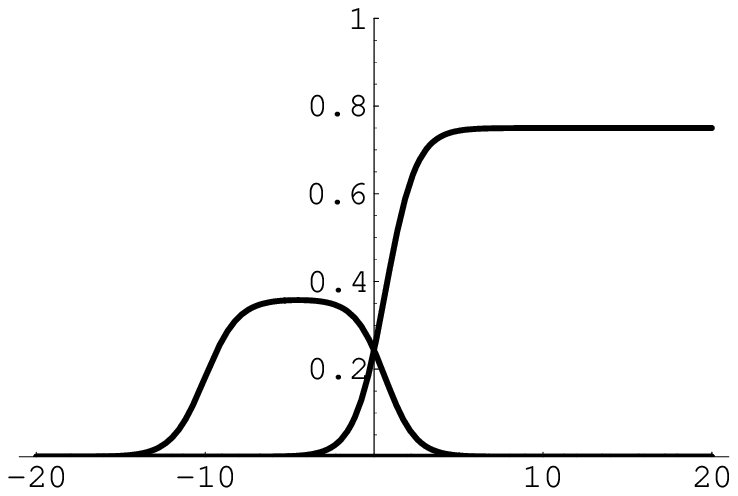}
\hspace{0.6cm}
\includegraphics[height=3cm]{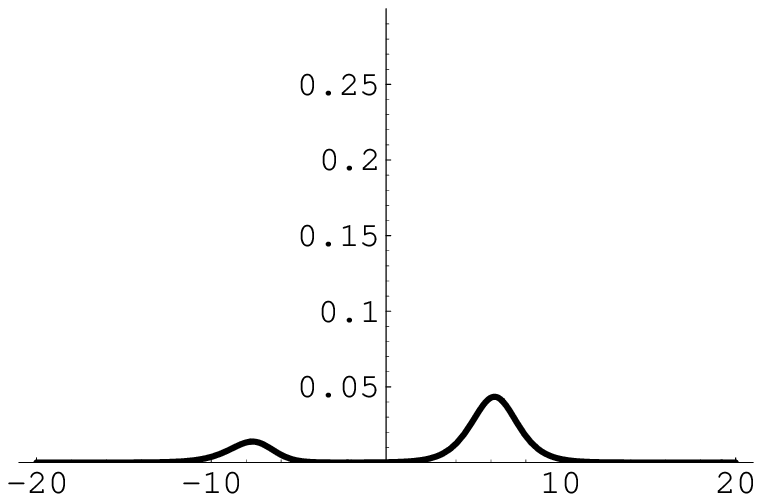} \hspace{0.6cm}
 \includegraphics[height=3cm]{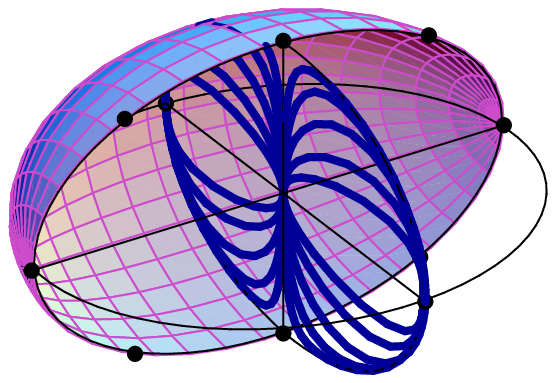}}
\caption{\small \textit{Kink form factor (a), energy density (b)
and orbits (c) for $K_2^{OC}(b)$ kinks. }}
\end{figure}
One of the non-null components, $\phi_2$, is again of kink form
but the other, $\phi_3$ is bell-shaped. The energy density is
localized at two different points, i.e. this kind of kinks are
formed by two basic lumps and are two-body solitary waves. They
are made from one $K_1^{OB}$ and one $K_1^{BC}$ kink, although due
to the nonlinearity of (\ref{eq:k2bc}) the composition is much
more intricate than the simple sum of expressions (\ref{eq:kob})
and (\ref{eq:kbc}). The value of $b$ determines the distance
between these lumps; if $b=0$, the centers of the one-body
solitary waves coincide  \cite{modeloa} and they are superposed.
Moreover,
\[
E[K_2^{OC}(b)]= E(K_1^{OB}) +
E(K_1^{BC})=\frac{\bar\sigma_2^4}{2\sqrt{2}} \qquad .
\]
The energy depends on the eccentricity of the ellipse $e_3$. We
remark that the fact of finding the combination of one $K_1^{OB}$
and one $K_1^{BC}$ as static solutions of (\ref{eq:partial}) means
that there are no forces between these basic particles when they
are at rest or their relative velocity is zero. The singular
member $K_2^{OC}(0)$ is located on the $\phi_2$ axis between the
points $C_+$ and $C_-$.

\textit{2. $K_2^{BD}(b)$:} On the ellipsoid $E \equiv \phi_1^2
+\frac{\phi_2^2}{\bar\sigma_2^2}+\frac{\phi_3^2}{\bar\sigma_3^2}=1$,
or $\lambda=0$, the Hamiltonian (\ref{eq:hamiltonian}) and the
Hamilton-Jacobi equation reads:
\begin{equation}
H=\frac{2 P_3(\mu)p_\mu^2 +P_4(\mu)}{\mu(\mu-\nu)}+ \frac{2
P_3(\nu)p_\nu^2 +P_4(\nu)}{\nu(\nu-\mu)}\hspace{1cm} ,
\hspace{1cm} \frac{\partial {\cal S}}{\partial x}+H\left(
\frac{\partial {\cal S}}{\partial \mu},\frac{\partial {\cal
S}}{\partial \nu},\mu,\nu \right)=0 \label{eq:hamilton2}\qquad .
\end{equation}
Separation of variables ${\cal S}={\cal S}_x(x)+{\cal
S}_\lambda(\lambda)+{\cal S}_\mu(\mu)$ affords the solution of
(\ref{eq:hamilton2}) for the Hamilton principal function via
quadratures: $S_x=-E x$,
\[
{\cal S}_\mu={\rm s}_\mu \int d\mu
\sqrt{\frac{(F+\frac{1}{2}E\mu)\mu}{(1-\mu)(\bar\sigma_2^2-\mu)(\bar\sigma_3^2-\mu)}
+\frac{\mu^2}{2}};\hspace{0.5cm} {\cal S}_\nu={\rm s}_\nu \int
d\nu
\sqrt{\frac{(F+\frac{1}{2}E\nu)\nu}{(1-\nu)(\bar\sigma_2^2-\nu)(\bar\sigma_3^2-\nu)}
+\frac{\nu^2}{2}} \quad ,
\]
if $E$ and $F$ are the separation constants. The finite action
trajectories, $E=F=0$, are determined by:
\begin{enumerate}
\item Orbits
\[
\gamma_2=\frac{\partial {\cal S}}{\partial F} \hspace{0.5cm}
\Leftrightarrow \hspace{0.5cm} e^{\sqrt{2}\sigma_{32}^2
\gamma_2}=\left[
\frac{(\bar\sigma_2^2-\mu)^\frac{1}{\sigma_2^2}}{(1-\mu)^{\frac{1}{\sigma_2^2}
-\frac{1}{\sigma_3^2}}(\mu-\bar\sigma_3^2)^\frac{1}{\sigma_3^2 }}
\right]^{{\rm s}_\mu} \left[
\frac{(\nu-\bar\sigma_2^2)^\frac{1}{\sigma_2^2}}{(1-\nu)^{\frac{1}{\sigma_2^2}
-\frac{1}{\sigma_3^2}}(\nu-\bar\sigma_3^2)^\frac{1}{\sigma_3^2 }}
\right]^{{\rm s}_\nu}
\]
\item Time schedules
\[
\gamma_1=\frac{\partial {\cal S}}{\partial E}\hspace{0.5cm}
\Leftrightarrow \hspace{0.5cm} e^{2 \sqrt{2}\sigma_{32}^2
\bar{x}}= \left[
\frac{(\bar\sigma_2^2-\mu)^\frac{1-\sigma_2^2}{\sigma_2^2}}{(1-\mu)^{\frac{1}{\sigma_2^2}
-\frac{1}{\sigma_3^2}}(\mu-\bar\sigma_3^2)^\frac{1-\sigma_3^2}{\sigma_3^2
}} \right]^{{\rm s}_\mu} \left[
\frac{(\nu-\bar\sigma_2^2)^\frac{1-\sigma_2^2}{\sigma_2^2}}{(1-\nu)^{\frac{1}{\sigma_2^2}
-\frac{1}{\sigma_3^2}}(\nu-\bar\sigma_3^2)^\frac{1-\sigma_3^2}{\sigma_3^2
}} \right]^{{\rm s}_\nu}
\]
In the field theoretical problem ,the time schedule becomes the
kink form factor.
\end{enumerate}

It is convenient to rearrange the above expressions and write:
\begin{equation}
e^{\sqrt{2} \sigma_2^2 (\bar\sigma_3^2 \gamma_2-2\bar{x})}=\left[
\frac{1-\mu}{\bar\sigma_2^2-\mu} \right]^{{\rm s}_\mu} \left[
\frac{1-\nu}{\nu-\bar\sigma_2^2} \right]^{{\rm s}_\nu} \!\!;
\hspace{0.3cm} e^{\sqrt{2} \sigma_3^2 (\bar\sigma_2^2
\gamma_2-2\bar{x})}=\left[ \frac{1-\mu}{\mu-\bar\sigma_3^2}
\right]^{{\rm s}_\mu} \left[ \frac{1-\nu}{\nu-\bar\sigma_3^2}
\right]^{{\rm s}_\nu} \label{eq:nu1} \qquad .
\end{equation}
Choosing ${\rm s}_\mu={\rm s}_\nu$ in (\ref{eq:nu1}), solitary
waves living in in the topological sector ${\cal C}^{OC}$ are
obtained. Back in Cartesian coordinates we find:
\begin{eqnarray}
\bar\phi_1(x;b)&=&
\frac{(-1)^\alpha}{\sqrt{1+\frac{\sigma_2^2}{\sigma_{32}^2}\,
e^{2\sqrt{2} \sigma_3^2 x}+\frac{\sigma_3^2}{\sigma_{32}^2}
\,b^2\, e^{2 \sqrt{2} \sigma_2^2 x} }} \nonumber
\\
\bar\phi_2(x;b)&=& \frac{b}{\sqrt{b^2 +
\frac{\sigma_2^2}{\sigma_3^2} \,e^{2 \sqrt{2}\sigma_{32}^2
x}+\frac{\sigma_{32}^2}{\sigma_3^2}\,
e^{-2\sqrt{2}\sigma_2^2 x}}} \label{eq:tk2db} \\
\bar\phi_3(x;b)&=& \frac{(-1)^\beta}{\sqrt{1+
\frac{\sigma_{32}^2}{\sigma_2^2}\, e^{-2\sqrt{2}\sigma_{3}^2x}+
\frac{\sigma_{3}^2}{\sigma_2^2} \,b^2 \,e^{-2\sqrt{2}\sigma_{32}^2
x}}} \nonumber \qquad .
\end{eqnarray}
$\alpha,\beta=0,1$ determine the quadrant of the ellipsoid
delimited by the $\phi_1=0$ and $\phi_3=0$ planes where the
solutions are confined, see Figure 6(c). Again, this is a
one-parametric family of solitary waves, denoted as $K_2^{DB}(b)$.
An $\alpha=\beta=0$ member of this family is depicted in Figure
6(c), asymptotically connecting the points $D_+$ and $B_+$.
\begin{figure}[hbt]
\centerline{\includegraphics[height=3cm]{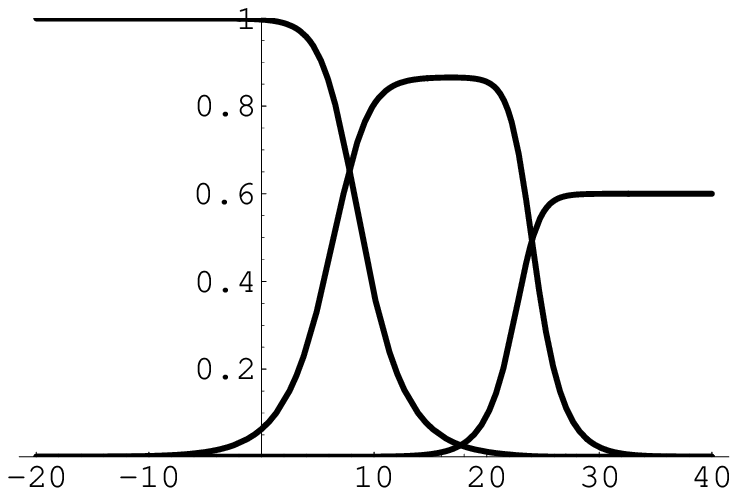}
\hspace{0.6cm}
\includegraphics[height=3cm]{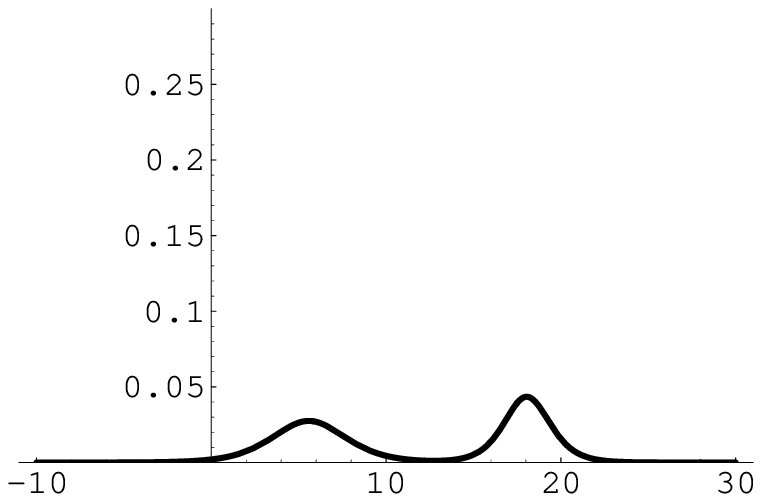} \hspace{0.6cm}
 \includegraphics[height=3cm]{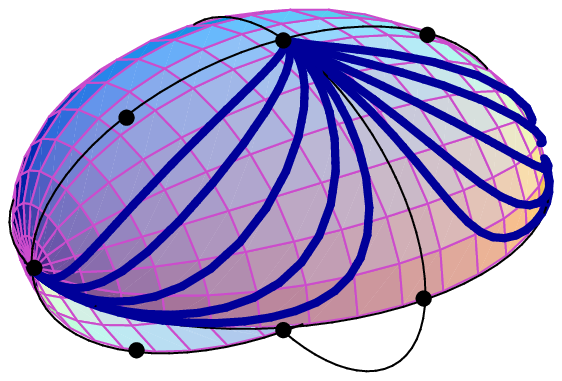}}
\caption{\small \textit{Form factor (a), energy density (b) and
orbits (c) for  $K_2^{BD}$ kinks. }}
\end{figure}
Note that these solitary waves have two kink-shaped non-null
components and the third one is bell-shaped. The energy density,
however, shows two basic lumps, one $K_1^{BC}$ and one $K_1^{CD}$,
a composition pointing to lack of interaction between these
particles at rest. The energy of these two-body kinks depends on
the eccentricity of the ellipse $e_2$,
\[
E[K_2^{BD}(b)]= E(K_1^{BC}) + E(K_1^{CD})=\frac{\sigma_3^2
(2-\sigma_3^2)}{2\sqrt{2}}
\]
The $K_2^{BD}(0)$ kink orbit is the ellipse $e_2$; for $b=0$
$E(K_1^{BC})$ and $E(K_1^{CD})$ are superposed, see
\cite{modeloa}.

In sum, there are two kinds of static two-body solitary waves
formed by either a $K_1^{OB}$ plus a $K_1^{BC}$ kinks or a
$K_1^{BC}$ plus $K_1^{CD}$ lumps. We remark that there is no
static configurations involving a $K_1^{OB}$ and $K_1^{CD}$ kinks
or two particles of the same kind. Thus, these particles interact
one with each other.

\subsection{Three-body solitary waves}

\textit{1. $K_3^{OD}(a,b)$:}The Hamilton-Jacobi equation
\[
\frac{\partial {\cal S}}{\partial x}+H\left(\frac{\partial {\cal
S}}{\partial \lambda},\frac{\partial {\cal S}}{\partial
\mu},\frac{\partial {\cal S}}{\partial \nu},\lambda,\mu,\nu
\right)=0
\]
for the complete Hamiltonian (\ref{eq:hamiltonian}) of the
analogous mechanical system is in any case separable. Writing the
Hamilton principal function in the form ${\cal S}=-Ex+{\cal
S}_\lambda+{\cal S}_\mu+{\cal S}_\nu$, one is led to the
quadratures:
\begin{eqnarray*}
{\cal S}_\lambda &=& {\rm s}_\lambda \int \sqrt{\frac{E\lambda^2
+F_1 \lambda
+F_3}{2(1-\lambda)(\bar\sigma_2^2-\lambda)(\bar\sigma_3^2-\lambda)}+
\frac{\lambda^2}{2}}\,\, d\lambda \\
{\cal S}_\mu &=&  {\rm s}_\mu \int \sqrt{\frac{E\mu^2 +F_1 \mu
+F_2}{2(1-\mu)(\bar\sigma_2^2-\mu)(\bar\sigma_3^2-\mu)}+
\frac{\mu^2}{2}}\,\, d\mu \\
{\cal S}_\nu &=&  {\rm s}_\nu \int \sqrt{\frac{E\nu^2 +F_1 \nu
+F_2}{2(1-\nu)(\bar\sigma_2^2-\nu)(\bar\sigma_3^2-\nu)}+
\frac{\nu^2}{2}}\,\, d\nu \qquad \qquad .
\end{eqnarray*}
Again, finite action trajectories $E=F_1=F_2=0$ are determined by:

\vspace{0.2cm}

\textbf{1.} Orbits, given by the intersection of two surfaces in
$P(0)$.

\vspace{0.15cm}

\textbf{1a.} $\gamma_2=\frac{\partial {\cal S}}{\partial F_1}$,
or, {\small
\[e^{2\sqrt{2}\sigma_{32}^2 \gamma_2}=\left[\frac{
(\bar\sigma_2^2-\lambda)^\frac{1}{\sigma_2^2}}{
(1-\lambda)^{\frac{\sigma_{32}^2}{\sigma_2^2\sigma_3^2}}
(\bar\sigma_3^2-\lambda)^\frac{1}{\sigma_3^2} }\right]^{{\rm
s}_\lambda} \left[\frac{
(\bar\sigma_2^2-\mu)^\frac{1}{\sigma_2^2}}{
(1-\mu)^{\frac{\sigma_{32}^2}{\sigma_2^2\sigma_3^2}}
(\mu-\bar\sigma_3^2)^\frac{1}{\sigma_3^2} }\right]^{{\rm s}_\mu}
\left[\frac{ (\nu-\bar\sigma_2^2)^\frac{1}{\sigma_2^2}}{
(1-\nu)^{\frac{\sigma_{32}^2}{\sigma_2^2\sigma_3^2}}
(\nu-\bar\sigma_3^2)^\frac{1}{\sigma_3^2} }\right]^{{\rm s}_\nu}
\]}

\vspace{0.15cm}

\textbf{1b.} $\gamma_3=\frac{\partial {\cal S}}{\partial F_2}$,
or, {\small
\[e^{2\sqrt{2}\sigma_{32}^2 \gamma_3}=\left[\frac{
\lambda^{\frac{\sigma_{32}^2}{\bar\sigma_2^2 \bar\sigma_3^2}}
(\bar\sigma_2^2-\lambda)^\frac{1}{\sigma_2^2\bar\sigma_2^2}}{
(1-\lambda)^{\frac{\sigma_{32}^2}{\sigma_2^2\sigma_3^2}}
(\bar\sigma_3^2-\lambda)^\frac{1}{\sigma_3^2\bar\sigma_3^2}
}\right]^{{\rm s}_\lambda}\left[\frac{
\mu^{\frac{\sigma_{32}^2}{\bar\sigma_2^2 \bar\sigma_3^2}}
(\bar\sigma_2^2-\mu)^\frac{1}{\sigma_2^2\bar\sigma_2^2}}{
(1-\mu)^{\frac{\sigma_{32}^2}{\sigma_2^2\sigma_3^2}}
(\mu-\bar\sigma_3^2)^\frac{1}{\sigma_3^2\bar\sigma_3^2}
}\right]^{{\rm s}_\mu} \left[\frac{
\nu^{\frac{\sigma_{32}^2}{\bar\sigma_2^2 \bar\sigma_3^2}}
(\nu-\bar\sigma_2^2)^\frac{1}{\sigma_2^2\bar\sigma_2^2}}{
(1-\nu)^{\frac{\sigma_{32}^2}{\sigma_2^2\sigma_3^2}}
(\nu-\bar\sigma_3^2)^\frac{1}{\sigma_3^2\bar\sigma_3^2}
}\right]^{{\rm s}_\nu} \]}

\textbf{2.} Time schedules, or kink form factor:
$\gamma_1=\frac{\partial {\cal S}}{\partial E}$, or, {\small
\[
e^{2\sqrt{2}\sigma_{32}^2 \bar{x}}=\left[\frac{
(\bar\sigma_2^2-\lambda)^\frac{\bar\sigma_2^2}{\sigma_2^2}}{
(1-\lambda)^{\frac{\sigma_{32}^2}{\sigma_2^2\sigma_3^2}}
(\bar\sigma_3^2-\lambda)^\frac{\bar\sigma_3^2}{\sigma_3^2}
}\right]^{{\rm s}_\lambda} \left[\frac{
(\bar\sigma_2^2-\mu)^\frac{\bar\sigma_2^2}{\sigma_2^2}}{
(1-\mu)^{\frac{\sigma_{32}^2}{\sigma_2^2\sigma_3^2}}
(\mu-\bar\sigma_3^2)^\frac{\bar\sigma_3^2}{\sigma_3^2}
}\right]^{{\rm s}_\mu} \left[\frac{
(\nu-\bar\sigma_2^2)^\frac{\bar\sigma_2^2}{\sigma_2^2}}{
(1-\nu)^{\frac{\sigma_{32}^2}{\sigma_2^2\sigma_3^2}}
(\nu-\bar\sigma_3^2)^\frac{\bar\sigma_3^2}{\sigma_3^2}
}\right]^{{\rm s}_\nu} \]}

These expressions are conveniently rearranged in the simpler form:
\begin{eqnarray}
\left[\frac{\lambda}{1-\lambda}\right]^{{\rm s }_\lambda}
\left[\frac{\mu}{1-\mu}\right]^{{\rm s }_\mu} \left[\frac{
\nu}{1-\nu}\right]^{{\rm s }_\nu}
&=&e^{2\sqrt{2}(x-(\bar\sigma_2^2+\bar\sigma_3^2)
\gamma_2 +\bar\sigma_2^2 \bar\sigma_3^2 \gamma_3)} \nonumber \\
\left[\frac{\lambda }{\bar\sigma_2^2-\lambda}\right]^{{\rm s
}_\lambda} \left[\frac{\mu}{\bar\sigma_2^2-\mu}\right]^{{\rm s
}_\mu} \left[\frac{\nu}{\nu-\bar\sigma_2^2}\right]^{{\rm s
}_\nu}&=& e^{2\sqrt{2}\bar\sigma_2^2 (x-(2-\bar\sigma_3^2)
\gamma_2 +\bar\sigma_3^2 \gamma_3)}  \label{eq:nu3} \\
\left[\frac{\lambda }{\bar\sigma_3^2-\lambda}\right]^{{\rm s
}_\lambda} \left[\frac{\mu}{\mu-\bar\sigma_3^2}\right]^{{\rm s
}_\mu} \left[\frac{\nu}{\nu-\bar\sigma_3^2}\right]^{{\rm s
}_\nu}&=& e^{2\sqrt{2}\bar\sigma_3^2 (x-(2-\bar\sigma_2^2)
\gamma_2 +\bar\sigma_2^2 \gamma_3)} \nonumber
\end{eqnarray}

Choosing ${\rm s}_\lambda={\rm s}_\mu={\rm s}_\nu=1$ in
(\ref{eq:nu3}) we obtain, back in Cartesian coordinates, the
two-parametric family of solitary waves, denoted as
$K_3^{OD}(a,b)$:
\begin{eqnarray} \bar\phi_1 (x;a,b) &=& \frac{(-1)^\alpha}{\sqrt{1+\frac{\sigma_2^2 \sigma_3^2}{
\bar\sigma_2^2 \bar\sigma_3^2}\,e^{2\sqrt{2}\bar{x}}
+\frac{\sigma_2^2}{\bar\sigma_3^2}\, b^2 e^{2\sqrt{2}
\sigma_3^2 \bar{x}}+\frac{\sigma_3^2}{\bar\sigma_2^2}\,a^2 e^{2\sqrt{2} \sigma_2^2 \bar{x}}  }} \nonumber \\
\bar\phi_2 (x;a,b) &=&
\frac{a}{\sqrt{\frac{a^2}{\bar\sigma_2^2}+\frac{\sigma_2^2}{
\bar\sigma_2^2 \bar\sigma_3^2}\,e^{2\sqrt{2}\bar\sigma_2^2
\bar{x}} +\frac{1}{\sigma_3^2}\, e^{-2\sqrt{2} \sigma_2^2
\bar{x}}+\frac{\sigma_2^2}{\sigma_3^2 \bar\sigma_3^2}\, b^2
e^{2\sqrt{2}
\sigma_{32}^2 \bar{x}}  }} \label{eq:tk3od} \\
\bar\phi_3 (x;a,b) &=&
\frac{b}{\sqrt{\frac{b^2}{\bar\sigma_3^2}+\frac{\sigma_3^2}{
\bar\sigma_2^2 \bar\sigma_3^2}\, e^{2\sqrt{2}\bar\sigma_3^2
\bar{x}} +\frac{1}{\sigma_2^2} \,e^{-2\sqrt{2} \sigma_3^2
\bar{x}}+\frac{\sigma_3^2}{\sigma_2^2 \bar\sigma_2^2}\, a^2
e^{-2\sqrt{2} \sigma_{32}^2 \bar{x}}  }}\nonumber
\end{eqnarray}

Without counting the center of the kinks, related to $\gamma_1$,
these kinks depend generically on two parameters $a$ and $b$,
defined respectively in terms of $\gamma_2$ and $\gamma_3$ as:
$\sigma_{32}^2 a^2= e^{-2\sqrt{2}\sigma_2^2 \sigma_{32}^2
\gamma_2}$, $\sigma_{32}^2 b^2= e^{-2\sqrt{2}\sigma_3^2
\sigma_{32}^2 \gamma_3}$. Additionally, $\alpha=0 \, \, {\rm or}
\, \, 1$ specifies in which semi-space $\phi_1>0$ or $\phi_1<0$
these solutions live, see Figure 7(c). In Figure 7(a) a $K_3^{OD}$
kink orbit is plotted for $\alpha=0$. The particle leaves the
point $D_+$ at $x=-\infty$  and arrives in $O$ at $x=+\infty$,
tracing a orbit confined to be in a octant of the interior of the
ellipsoid $E$. The associated solitary wave thus lives in the
${\cal C}^{OD_+}$ sector of the configuration space and has one
kink-like and two bell-shaped components. The energy density of a
solitary wave like this is composed of three lumps: the basic
particles $K_1^{OB}$, $K_1^{BC}$ and $K_1^{CD}$, see Figure 7(b).

The energy of any kink solution of the (\ref{eq:tk3od}) family is
topological and does not depend neither on the $(a,b)$ parameters
nor on the coupling constants:
\begin{eqnarray*}
E[K_3^{OD}(a,b)]&=&|W(\bar{\phi}_1(\infty;a,b),
\bar{\phi}_2(\infty;a,b),
\bar{\phi}_3(\infty;a,b))-W(\bar{\phi}_1(-\infty;a,b),
\bar{\phi}_2(-\infty;a,b), \bar{\phi}_3(-\infty;a,b))|\\&=&
E(K_1^{OB}) + E(K_1^{BC}) +E(K_1^{CD})
=\frac{1}{2\sqrt{2}}\label{eq:ksr}
\end{eqnarray*}
shows degeneracy in energy and a curious kink energy sum rule.
\begin{figure}[hbt]
\centerline{\includegraphics[height=3cm]{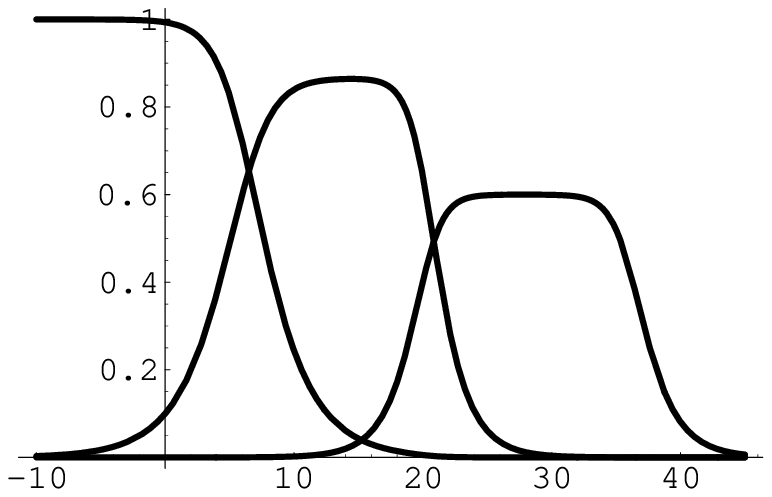}
\hspace{0.6cm}
\includegraphics[height=3cm]{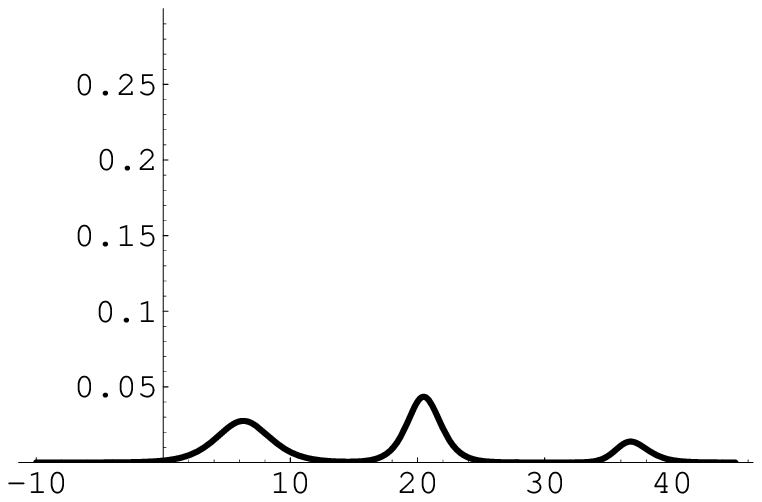} \hspace{0.6cm}
 \includegraphics[height=3cm]{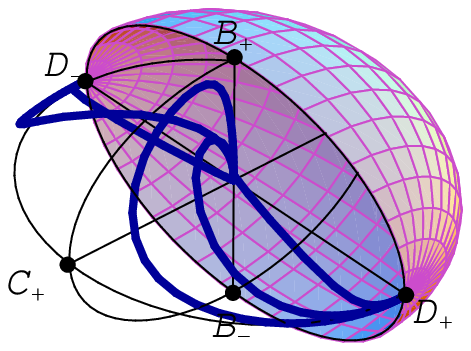}}
\caption{\small \textit{Kink form factor (a), energy density (b),
and orbits (c) for  $K_3^{OD}(a,b)$ kinks. }}
\end{figure}

The choice of the parameters $a$ and $b$ to draw Figure 7 aims at
identifying the basic components of any $K_3^{OD}(a,b)$ kink. In
general, the parameter $a$ is set in order to find kinks in the
$\phi_3=0$ plane for $a=0$, whereas the parameter $b$ runs through
$\mathbb{R}$. In Figure 8(c) several orbits of the $K_3^{OD}(0,b)$
kink sub-family are shown. Apparently these solutions are made of
two lumps, as illustrated in Figure 8(b). The paradox is
ficticious because the $K_1^{BC}$ and $K_1^{OB}$ lumps are
superposed for all the members of this sub-family. Therefore,
these kinks also involve the three basic particles and we
recognize $a$ as the measure of the distance between $K_1^{BC}$
and $K_1^{OB}$ .

\begin{figure}[hbt]
\centerline{\includegraphics[height=3cm]{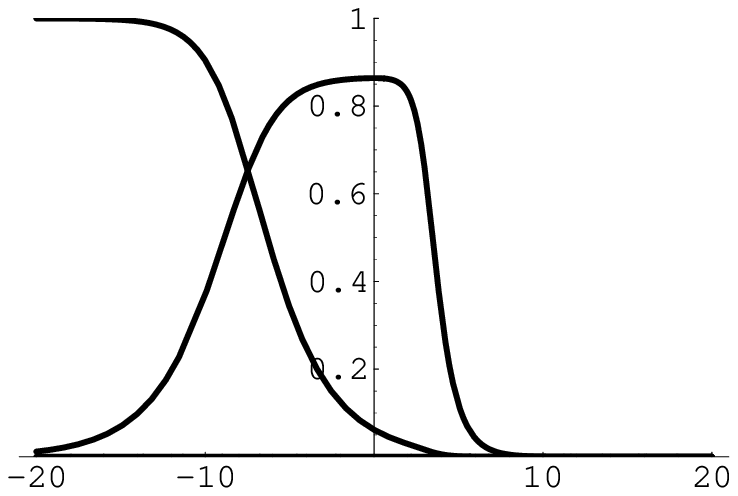}
\hspace{0.6cm}
\includegraphics[height=3cm]{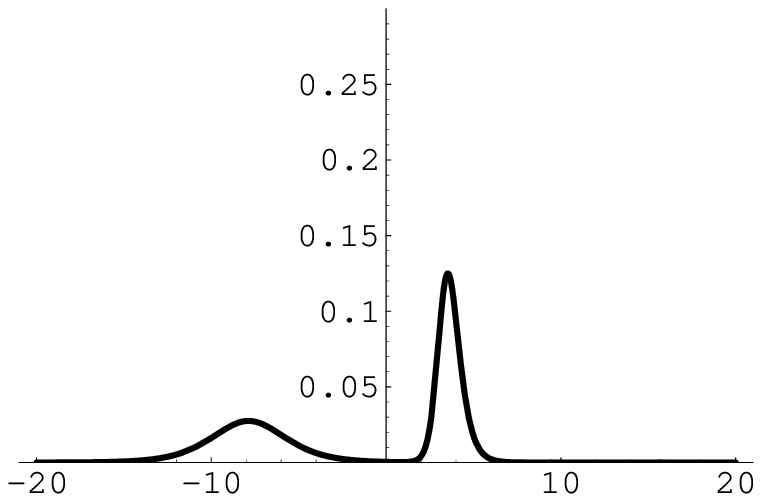} \hspace{0.6cm}
 \includegraphics[height=3cm]{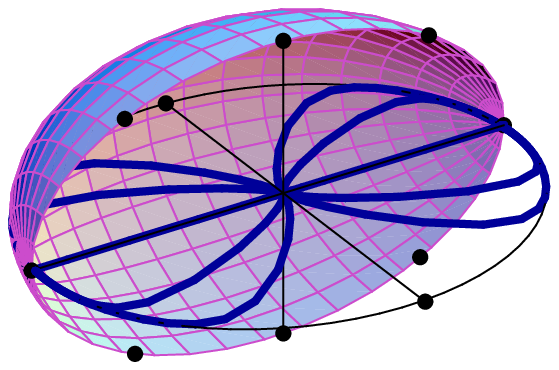}}
\caption{\small \textit{Form factor (a), energy density (b) and
orbits (c) for $K_3^{OD}(0,b)$ kinks. }}
\end{figure}
In the same vein, $K_3^{OD}(a,0)$ solitary waves live in the plane
$\phi_2=0$. Each member of the $b=0$ family displays two lumps,
one $K_1^{OB}$ kink and a superposition of a $K_1^{BC}$ and a
$K_1^{CD}$ basic particles. $b$ thus measures the distance between
the $K_1^{BC}$ and $K_1^{CD}$ basic kinks.  There exists a member
of the family in the intersection of the two sub-families, i.e.,
the $K_3^{OD}(0,0)$ kink, whose orbit lies on the $\phi_1$ axis.
Generically, the $K_3(a,b)$ solitary waves are three-body kinks,
but in the $K_3(0,0)$ configuration the three particles or lumps
are completely fused. In Figure 9 the energy density of
$K_3(10,b)$ solitary waves is depicted for several values of the
parameter $b$. A similar pattern would be observed by setting the
value of $b$ to be constant and letting the value of $a$ vary,
although the r$\hat{\rm o}$le of the basic particles would be
swapped.
\begin{figure}[hbt]
\centerline{\includegraphics[height=2cm]{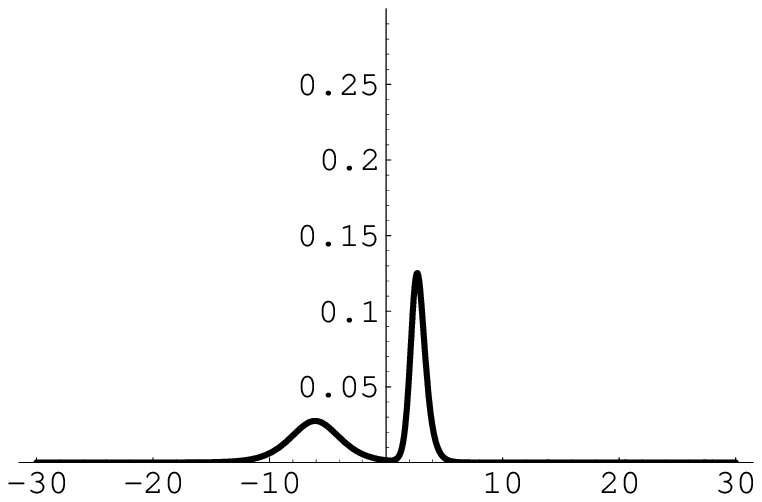}
\includegraphics[height=2cm]{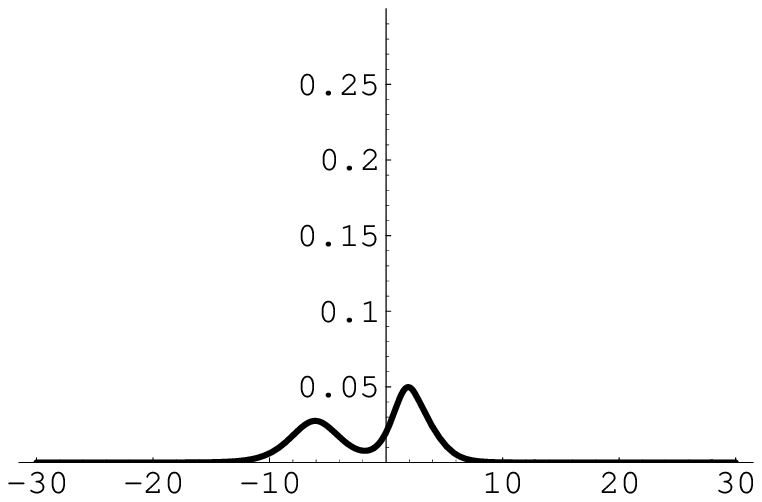}
\includegraphics[height=2cm]{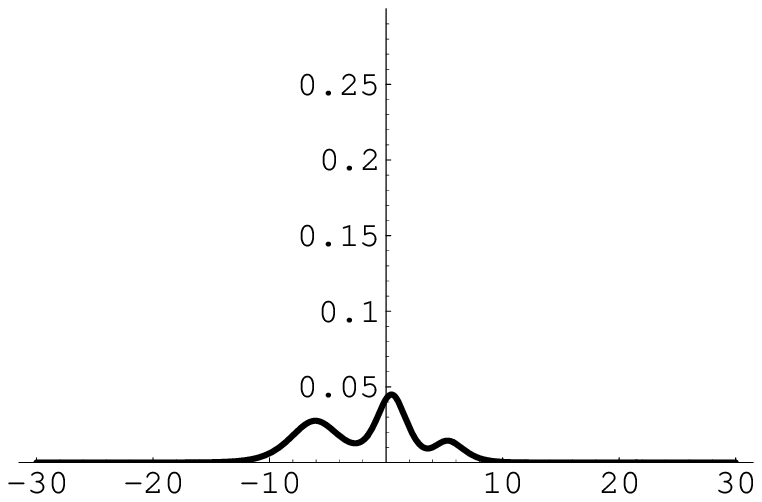}
\includegraphics[height=2cm]{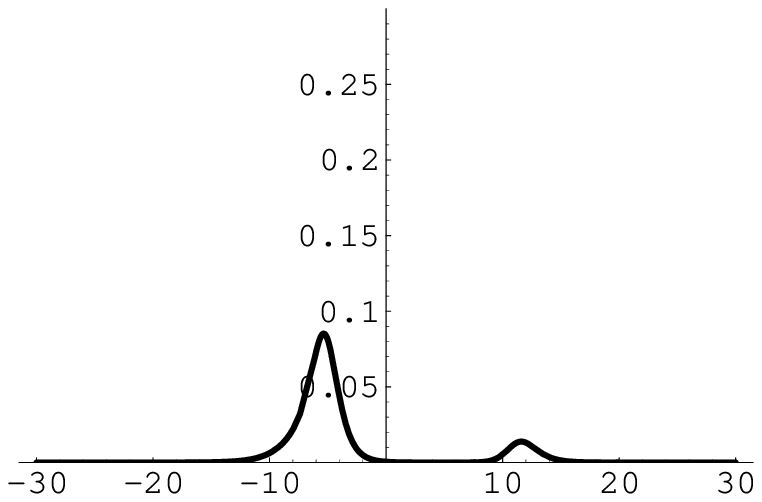}
\includegraphics[height=2cm]{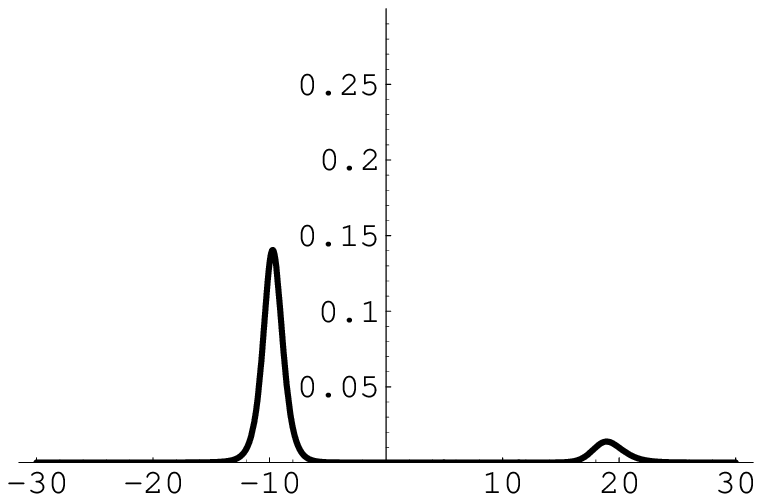}}
\caption{\small \textit{Energy Density for $K_3^{OD}(10,b)$ kinks:
a) $b=0$, Green zone B, b) $b=5$, Green zone B, c) $b=10$, Blue
zone D, d) $b=240$, Green zone C, and e) $b=10000$, Green zone C.
All the colored zones are shown in Figure 10. }}
\end{figure}

It is interesting to remark that the choice of signs, providing
the solutions (\ref{eq:tk3od}), is such that the inverse image of
\[
s_\lambda S_\lambda(\lambda)+s_\mu S_\mu(\mu)+s_\nu S_\nu(\nu)
\]
in the coordinate transformation (\ref{eq:elipticas}) is the
Hamilton characteristic function in ${\mathbb R}^3$:
\begin{equation}
W(\phi_1,\phi_2,\phi_3)=\frac{1}{\sqrt{8}} \left[
(\phi_1^2+\phi_2^2+\phi_3^2-1)^2+ 2\sigma_2^2 \phi_2^2 + 2
\sigma_3^2 \phi_3^2 \right] \qquad . \label{eq:superpot}
\end{equation}
Thus,
\begin{enumerate}

\item (\ref{eq:tk3od}) are the gradient flow lines
of $W(\phi_1,\phi_2,\phi_3)$:
\begin{equation}
\frac{d\phi_1}{d x }=\mp \frac{\partial
W}{\partial\phi_1}\hspace{1cm} , \hspace{1cm} \frac{d\phi_2}{d x
}=\mp \frac{\partial W}{\partial\phi_2} \hspace{1cm} ,
\hspace{1cm} \frac{d\phi_3}{d x }=\mp \frac{\partial
W}{\partial\phi_3} \qquad . \label{eq:gfl}
\end{equation}
The ODE system (\ref{eq:gfl}) is separable and tantamount to
(\ref{eq:nu3}) in elliptic coordinates.

\item $W(\phi_1,\phi_2,\phi_3)$ is no more than the potential
energy of a very well known integrable system: the Garnier system,
the analogous mechanical system of the field theoretical model
discussed in \cite{Aai3} and \cite{Aai4}. Our model is chosen in
such a way that the potential energy of the mechanical analogous
system is the square of the norm of the gradient of the potential
energy of the Garnier system. No wonder integrability is found!.

\end{enumerate}

In sum, there exist three-body solitary wave solutions that are
static configurations formed by the three kinds of basic
particles, with no forces between lumps.

\subsubsection{The moduli space of three-body kinks}

\begin{figure}[hbt]
\centerline{\includegraphics[height=5cm]{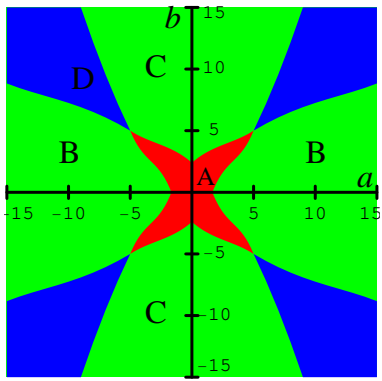}}
\caption{\small \textit{Tiling of the moduli space according to
the number of lumps. }}
\end{figure}

To make these statements more precise, we need a better
understanding of the physical meaning of the $(a,b)$ coordinates
parametrizing the moduli space of $K_3$ solitary waves. To achieve
this goal, the number of lumps should be identified with the
number of maxima of the energy density of the $K_3$ solution on
the line. For instance, in the
$\sigma_3^2=2\sigma_2^2=\frac{1}{3}$ case, the energy density
reads: {\footnotesize\[ {\cal E}_{\rm K}(z)=\frac{2z(2 a^2+(3a^4+4
b^2)z+12 a^2 b^2 z^2+(10 a^2+a^4 b^2+12 b^4)z^3+(32 b^2+2 a^2
b^4)z^4+(27 +6 a^2 b^2)z^5+8 a^2 z^6+b^2 z^7)}{27(1+a^2 z+b^2
z^2+z^3)^3 },
\]}
where $z=e^{\frac{2\sqrt{2}x}{3}}$. It is not possible to find
analytically the extremal points of ${\cal E}_{\rm K}(z)$ because
one should find the roots of a sixth order polynomial as a
function of $a$ and $b$. Solutions of the equation
\begin{equation}
\frac{\partial {\cal E}_{\rm K}}{\partial z}(z,a,b)=0 \qquad
\qquad , \label{eq:derden}
\end{equation}
however, can be studied numerically. We find the following
pattern in different ranges of $a$ and $b$:

\begin{itemize}
\item There is a single real solution of (\ref{eq:derden}) in the A
domain of the moduli space, see Figure 10. The solution is a
maximum of $\varepsilon_{\rm K}(z,a,b)$ and the three lumps are
glued together.
\item In the two B domains there are three real solutions to equation
(\ref{eq:derden}); two of them are maxima and the remaining one is
a minimum of $\varepsilon_{\rm K}(z,a,b)$. All over these regions
of the moduli space $K_1^{CD}$ and $K_1^{BC}$ sit on top of each
other, whereas $K_1^{OB}$ is apart.

\item Also in the two C domains there are three real solutions, two maxima
and one minimum, but now only $K_1^{CD}$ is
separated.

\item Finally, in the four D domains there are five real solutions of (\ref{eq:derden}),
three of which are maxima and the other two are minima of the
energy density, and the three basic kinks are totally split apart
from each other .
\end{itemize}

\subsection{Four-body solitary waves}

\textit{1. $K_4^{BB}(b)$:} The sign combinations ${\rm s}_\lambda
\neq {\rm s}_\mu$ in the (\ref{eq:nu0}) system of equations offer
a one-parametric family of solutions belonging to the ${\cal
C}^{BB}$ topological sectors and confined to living in the
$\phi_1=0$ plane. The Cartesian field profiles are:
\begin{eqnarray*}
\bar\phi_1(x)&=& 0 \\
\bar\phi_2(x)&=& \frac{(-1)^\alpha \bar\sigma_2^2 \sigma_{32}
(1+e^b e^{-2 \sqrt{2} \bar\sigma_3^2
\bar{x}})}{(\bar\sigma_2^2+\sigma_{32}^2 e^b
e^{-2\sqrt{2}\bar\sigma_3^2 \bar{x}}+\bar\sigma_3^2 e^b
e^{2\sqrt{2}\sigma_{32}^2 \bar{x}})^\frac{1}{2} (\bar\sigma_{32}^2
+\bar\sigma_2^2 e^b e^{-2\sqrt{2} \bar\sigma_3^2 x}+
\bar\sigma_3^2 e^{-2\sqrt{2} \bar\sigma_2^2 \bar{x} })^\frac{1}{2}} \\
\bar\phi_3(x)&=& \frac{\bar\sigma_3^2 \sigma_{32} (1- e^{-2
\sqrt{2} \bar\sigma_2^2 \bar{x}})}{(\bar\sigma_3^2+\sigma_{32}^2
e^{-2\sqrt{2}\bar\sigma_2^2 \bar{x}}+\bar\sigma_2^2 e^{-b}
e^{-2\sqrt{2}\sigma_{32}^2 \bar{x}})^\frac{1}{2}
(\bar\sigma_{32}^2 + \bar\sigma_2^2 e^b e^{-2\sqrt{2}
\bar\sigma_3^2 x}+ \bar\sigma_3^2 e^{-2\sqrt{2} \bar\sigma_2^2
\bar{x} })^\frac{1}{2}} \label{eq:k4cc} \qquad .
\end{eqnarray*}
Again, $b\in \mathbb{R}$ is a real parameter and $\alpha=0,1$. In
Figure 11(a) a $K_4^{BB}(b)$ kink is represented: the second
component is bell-shaped but $\bar{\phi}_3(x)$ is formed by two
kinks. The kink trajectory departs from the point $B_-$ and
arrives in $B_+$ when $x$ goes from $-\infty$ to $\infty$,
crossing through the focus $F_1$ of the ellipse $e_1$. Indeed,
this behavior is general and the trajectory of each member
$K_4^{BB}(b)$ passes through $F_1$ at the same \lq\lq time "
$\bar{x}=0$, see Figure 11(c). In Figure 11(b), the energy density
is plotted, showing the structure of this kind of solitary wave
solutions: these kinks are composed of three lumps; two of them
are easily recognizable as the basic particles $K_1^{OB}$ and
$K_1^{BC}$. The other kink is located between the previous ones
and has the same energy as the sum of the energy of the two basic
particles. Later stability analysis will show the lump in the
middle decaying to the $K_1^{OB}$ and $K_1^{BC}$ one-body kinks,
but exactly superposed in $K_2^{OC}(0)$ they are in unstable
equilibrium. In any case, the $K_4^{BB}(b)$ solutions are made of
four one-body kinks: two $K_1^{OB}$'s and two $K_1^{BC}$'s. The
energy of these solutions is:
\[
E[K_4^{BB}(b)]= E(K_1^{BO}) +  E[K_2^{OC}(0)] +  E(K_1^{BC}) =  2
E(K_1^{OB}) + 2 E(K_1^{BC})= \frac{\bar\sigma_2^4}{\sqrt{2}} \quad
.
\]
\begin{figure}[hbt]
\centerline{\includegraphics[height=3cm]{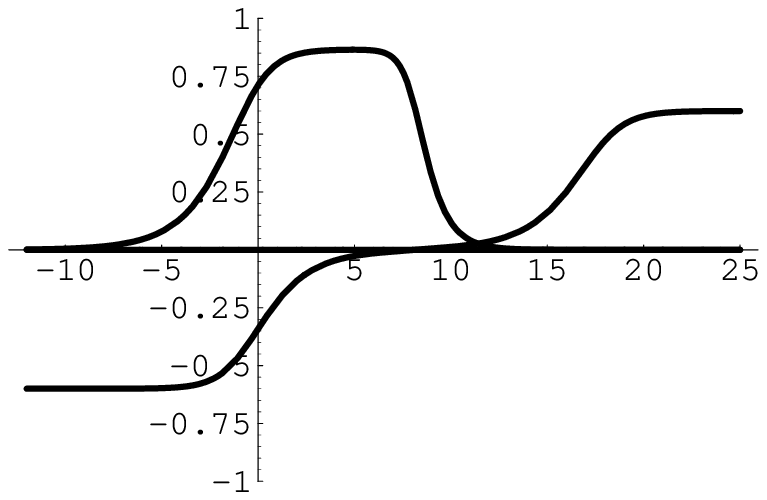}
\hspace{0.6cm}
\includegraphics[height=3cm]{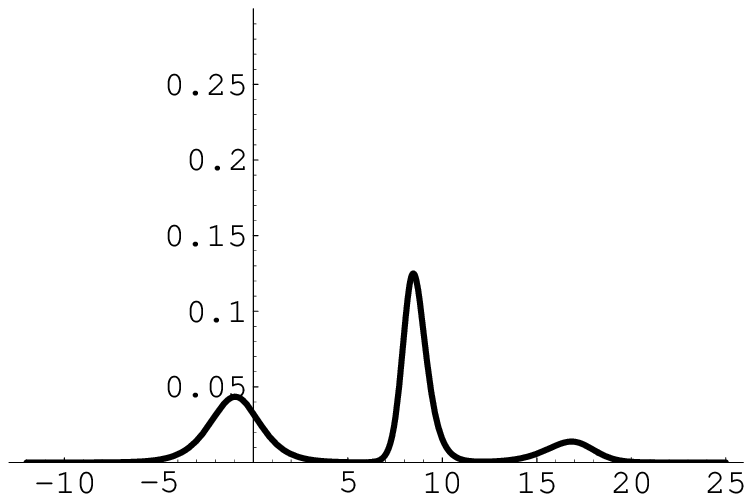} \hspace{0.6cm}
 \includegraphics[height=3cm]{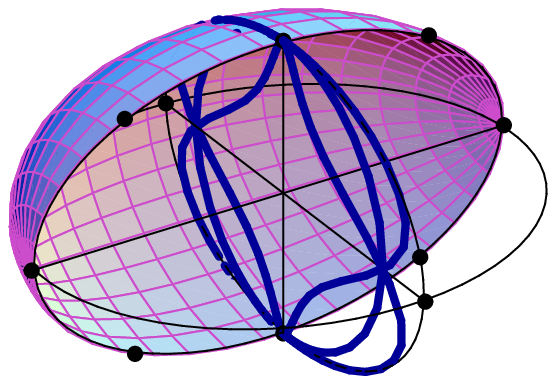}}
\caption{\small \textit{Kink form factor (a), energy density (b)
and orbits (c) for $K_4^{BB}$ kinks. }}
\end{figure}
An interesting point should be noticed: we claimed that the
presence of two particles of the same kind necessarily involves a
force between them. Here, we are dealing with a static
configuration of two pairs of basic particles. Thus, a delicate
balance between the $K_1^{OB}$-$K_1^{OB}$ and
$K_1^{BC}$-$K_1^{BC}$ forces must be reached. This is only
possible if one $K_1^{OB}$ and one $K_1^{BC}$ kink share exactly
their centers in between the other two basic particles, see Figure
11(b). Any small perturbation of this situation would release
forces, an argument for a dynamical explanation of instability.
The fate of these solutions as time passes is explained in
sub-Section 4.2 of Reference \cite{modeloa} for similar quadruple
solitary waves arising in that model.

\textit{2. $K_4^{CC}(b)$:} Another one-parametric family of kinks
exists on the ellipsoid ${\mathbb E}$ ($\lambda=0$). The sign
combinations ${\rm s}_\mu \neq {\rm s}_\nu$ in the system of
equations (\ref{eq:nu1}) provide the solutions in the original
fields:
\begin{eqnarray*}
\bar\phi_1(x)&=& \frac{(-1)^\alpha \sigma_2 \sigma_3 (1+e^b
e^{2\sqrt{2} \sigma_{32}^2 \bar{x}})}{\sqrt{(\sigma_2^2+
\sigma_{3}^2 e^b e^{2\sqrt{2}\sigma_{32}^2 \bar{x}} +\sigma_{32}^2
e^{2\sqrt{2} \sigma_3^2 \bar{x}}) (\sigma_3^2 +\sigma_2^2 e^b
e^{2\sqrt{2} \sigma_{32}^2 \bar{x}} + \sigma_{32}^2 e^b
e^{-2\sqrt{2} \sigma_2^2 \bar{x}})}
} \\
\bar\phi_2(x)&=&  \frac{\sigma_2 \bar\sigma_2 \sigma_{32} (-1+
e^{-2\sqrt{2} \sigma_{3}^2 \bar{x}})}{\sqrt{(\sigma_{32}^2+
\sigma_{2}^2 e^{-2\sqrt{2}\sigma_{3}^2 \bar{x}} +\sigma_{3}^2 e^b
e^{-2\sqrt{2} \sigma_2^2 \bar{x}}) (\sigma_2^2 +\sigma_3^2 e^{-b}
e^{-2\sqrt{2} \sigma_{32}^2 \bar{x}} + \sigma_{32}^2 e^{-2\sqrt{2}
\sigma_3^2 \bar{x}})}
} \\
\bar\phi_3(x)&=& \frac{(-1)^\beta \sigma_3 \bar\sigma_3
\sigma_{32} (1+e^b e^{-2\sqrt{2} \sigma_{2}^2
\bar{x}})}{\sqrt{(\sigma_{32}^2+ \sigma_{2}^2
e^{-2\sqrt{2}\sigma_{3}^2 \bar{x}} +\sigma_{3}^2 e^b e^{-2\sqrt{2}
\sigma_2^2 \bar{x}}) (\sigma_3^2 +\sigma_2^2 e^{b} e^{2\sqrt{2}
\sigma_{32}^2 \bar{x}} + \sigma_{32}^2 e^b e^{-2\sqrt{2}
\sigma_2^2 \bar{x}})} } \qquad \qquad .
 \label{eq:k4ccc}
\end{eqnarray*}
Two components are bell-shaped, $\phi_1$ and $\phi_3$, whereas the
third one, $\phi_2$,  is of two-kink form, see Figure 12(a).
$b=-\sqrt{2} \sigma_2^2 \sigma_{32}^2 \gamma_2$ is the parameter
that characterizes the family members; the values
$\alpha,\beta=0,1$ specify the quadrant of the ellipsoid ${\mathbb
E}$ in which the kink is located. These solitary waves belong to
the ${\cal C}^{CC}$ topological sectors and connect the points
$C^+$ and $C^-$ crossing the umbilicus points $A$, see Figure
12(c). In Figure 12(a), a $K_4^{CC}(b)$ kink is depicted leaving
from point $C_-$ and arriving at $C_+$ as $x$ goes from $-\infty$
to $\infty$. In Figure 12(b) the energy density is shown,
displayed by three lumps; two of them are easily identified as the
basic particles $K_1^{CD}$ and $K_1^{BC}$ but the more energetic
lump, located between the previous ones, corresponds to a
combination forming the $K_2^{DB}(0)$ kink. Thus, the
$K_4^{CC}(b)$ solutions are composites of four basic particles,
two $K_1^{CD}$ and two $K_1^{BC}$ ones. The energy depends only
the coupling constant $\sigma_3$:
\[
E[K_4^{CC}(b)]=E(K_1^{CB}) +  E[K_2^{DB}(0)] + E(K_1^{DC}) = 2
E(K_1^{CD}) + 2 E(K_1^{BC})=\frac{\sigma_3^2
(2-\sigma_3^2)}{\sqrt{2}} \,\, .
\]
\begin{figure}[hbt]
\centerline{\includegraphics[height=3cm]{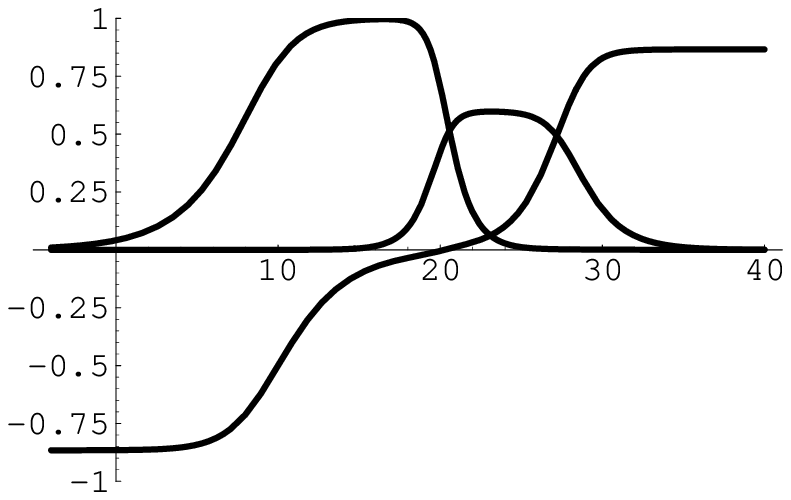}
\hspace{0.6cm}
\includegraphics[height=3cm]{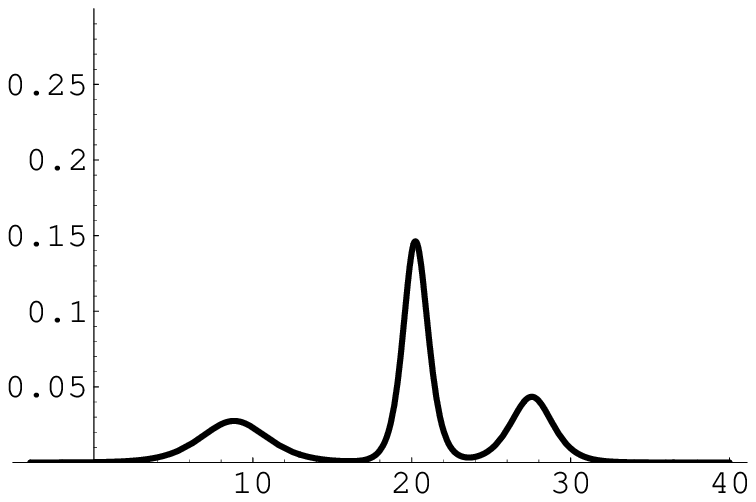} \hspace{0.6cm}
 \includegraphics[height=3cm]{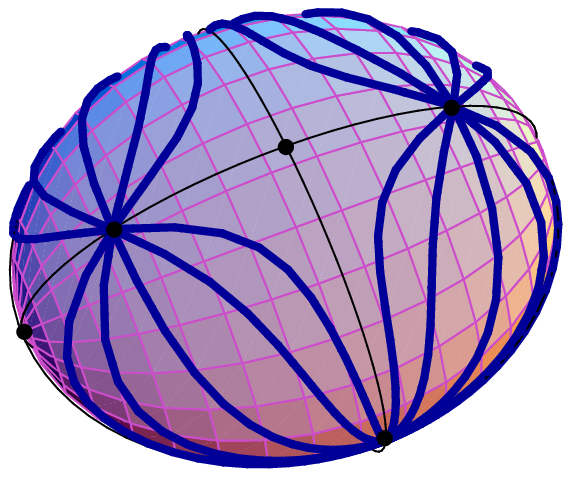}}
\caption{\small \textit{Kink form factor (a), energy density (b)
and orbits (c) for  $K_4^{CC}$ kinks. }}
\end{figure}
The $K_4^{CC}(b)$ four-body kinks are qualitatively identical to
the $K_4^{BB}(b)$ kinks replacing the ellipsoid ${\mathbb E}$ by
the $\phi_1=0$ plane ($\lambda=0$ and $\nu=1$ faces in the $P(0)$
cuboid). Thus, all considerations about stability and forces made
for the $K_4^{BB}(b)$ kinks also work for the $K_4^{CC}(b)$'s.

\subsection{Six-body solitary waves}

\textit{1. $K_6^{CC}(b)$:} There is a more complicated
one-parametric family of solitary waves in the ${\cal C}^{CC}$
topological sectors living in the $\phi_3=0$ plane. The potential
is restricted
\[
V(\phi_1,\phi_2,0)= (\phi_1^2+\phi_2^2)(\phi_1^2+\phi_2^2-1)^2+2
\sigma_2^2 \phi_2^2 (\phi_1^2+\phi_2^2-1)+\sigma_2^4 \phi_2^2
\]
to that discussed in the two scalar field model of Reference
\cite{modeloa}. In the present context, however, the Hamiltonian
(\ref{eq:hamiltonian}) reduces to either the I)
$\lambda=\bar{\sigma}_3^2$ or II) $\mu=\bar{\sigma}_3^2$  faces of
$P(0)$:
\[
H_I=\frac{H_\mu}{(\mu-\bar{\sigma}_3^2)(\mu-\nu)}+\frac{H_\nu}{(\nu-\bar{\sigma}_3^2)(\nu-\mu)}
\qquad , \qquad
H_{II}=\frac{H_\lambda}{(\lambda-\bar{\sigma}_3^2)(\lambda-\nu)}+\frac{H_\nu}{(\nu-\lambda)(\nu-\bar{\sigma}_3^2)}
\qquad .
\]
The Hamilton-Jacobi equations for $H_I$ and $H_{II}$ must be
solved consecutively and the two pieces should be continuously
glued afterwards, see Reference \cite{Aai3} for similar solutions
in a simpler model with a quartic potential energy. Sign
combinations for ${\rm s}_\lambda$, ${\rm s}_\mu$, ${\rm s}_\nu$
leading to these solitary waves are different from those giving
the $K_3^{OD}(0,b)$ kinks.

These solutions connect the points $C^+$ and $C^-$ and cross the
$F_2$ foci (see Figure 12(c)). The kink form factors are:
\begin{eqnarray*}
\bar\phi_1(x)&=& \frac{(-1)^\alpha \sigma_2 \left(1+e^{2 \sqrt{2}
\bar\sigma_2^2 \bar{x}}\right)}{\sqrt{\sigma_2^2+ \bar\sigma_2^2
e^{-b} e^{-2\sqrt{2} \bar x}+ e^{2\sqrt{2} \bar\sigma_2^2
 \bar{x}}} \sqrt{1+ \sigma_2^2
e^{2\sqrt{2}\bar\sigma_2^2 \bar
x}+ \bar\sigma_2^2 e^{b}e^{-2\sqrt{2}\sigma_2^2  \bar{x}}}} \\
\bar\phi_2(x)&=&\frac{\sigma_2 \bar\sigma_2^2 \left(e^{2\sqrt{2}
\bar{x}}-e^{b} \right) }{\sqrt{
 e^{2\sqrt{2}\sigma_2^2  \bar x}+ \sigma_2^2
e^{2\sqrt{2}  \bar{x}}+\bar\sigma_2^2 e^{b}} \sqrt{e^{b} e^{2
\sqrt{2} \bar\sigma_2^2  \bar
x}+ \sigma_2^2 e^{b}+\bar\sigma_2^2 e^{2\sqrt{2} \bar{x}}}} \\
\bar\phi_3(x)&=& 0 \label{eq:k4cc} \qquad \hspace{8cm} \qquad .
\end{eqnarray*}
In particular, a $K_6^{CC}(b)$ kink with $\alpha=0$ has been
depicted in Figure 12(a), departing from the point $C_-$ and
arriving in $C_+$ when $x$ goes from $-\infty$ to $\infty$. One
component is of bell form and the other is two kink-shaped. In
Figure 12(b) the $K_6^{CC}(b)$ energy density is shown.
Apparently, these solutions are formed by three lumps. One of them
is the basic particle $K_1^{CD}$; the more energetic lump in the
middle of the other two is the $K_3^{OD}(0)$ kink (the union of
the three basic particles), and the third lump is a $K_2^{OC}$
kink (superposition of the $K_1^{OB}$ and $K_1^{BC}$ basic kinks).
Therefore, the $K_6^{CC}(b)$ solutions involve six basic
particles, a pair of each type of basic lumps. These static
configurations describe an even more delicate equilibrium than
four-body lumps. The forces between basic kinks of the same type
are only balanced at the expense of joining them in a arrangement
such as that shown in Figure 12(b). Any relative displacement of
the centers of the composite particles would destroy these
unstable solutions. The energy of these six-body solitary waves
does not depend on the coupling constants:
\[
E[K_6^{CC}(b)]= E(K_1^{CD}) +  E[K_3^{DO}(0,0)] +  E[K_2^{OC}(0)]=
2 E(K_1^{CD}) + 2 E(K_1^{OB})+ 2 E(K_1^{BC})=\frac{1}{\sqrt{2}}
\qquad .
\]
\begin{figure}[hbt]
\centerline{\includegraphics[height=3cm]{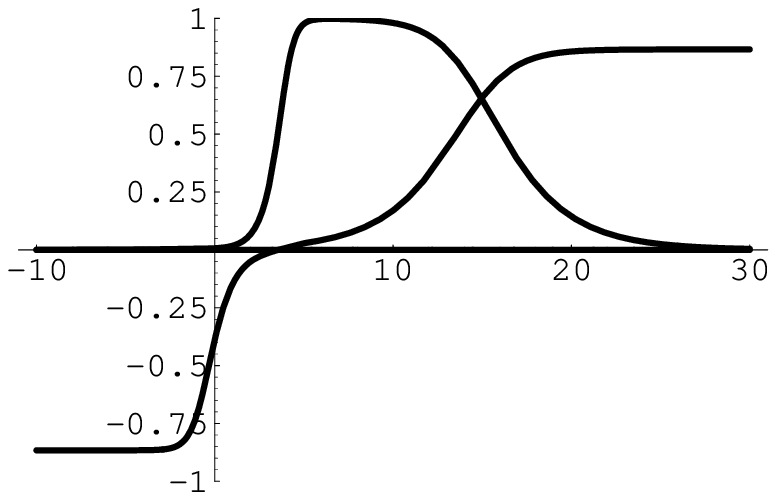}
\hspace{0.6cm}
\includegraphics[height=3cm]{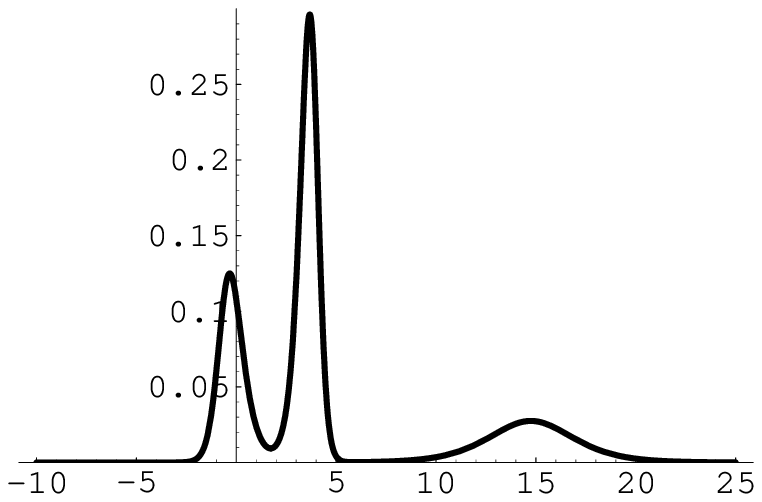} \hspace{0.6cm}
 \includegraphics[height=3cm]{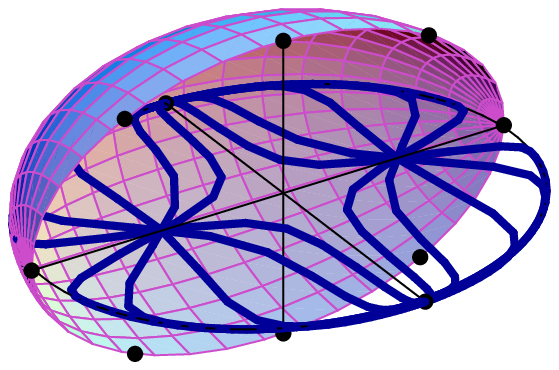}}
\caption{\small \textit{Kink form factor (a), energy density (b)
and orbits (c) for $K_6^{CC}$ kinks. }}
\end{figure}

\textit{2. $K_6^{BB}$:} Another one-parametric family of six-body
solitary waves exists in the ${\cal C}^{BB}$ topological sectors,
now located on the plane $\phi_2=0$. Also, the restricted
potential energy,
\[
V(\phi_1,0,\phi_2)= (\phi_1^2+\phi_3^2)(\phi_1^2+\phi_3^2-1)^2+2
\sigma_3^2 \phi_3^2(\phi_1^2+\phi_2^2+\phi_3^2-1)+\sigma_3^4
 \phi_3^2 \qquad ,
\]
corresponds to a two-component scalar field model studied in
\cite{modeloa}. There is one difference in this three-component
situation: as in the previous case, the Hamiltonian
(\ref{eq:hamiltonian}) reduces to either the I)
$\mu=\bar{\sigma}_2^2$ or II) $\nu=\bar{\sigma}_2^2$  faces of
$P(0)$:
\[
H_I=\frac{H_\lambda}{(\lambda-\bar{\sigma}_2^2)(\lambda-\nu)}+\frac{H_\nu}{(\nu-\lambda)(\nu-\bar{\sigma}_2^2)}
\qquad , \qquad
H_{II}=\frac{H_\lambda}{(\lambda-\mu)(\lambda-\bar{\sigma}_2^2)}+\frac{H_\mu}{(\mu-\lambda)(\mu-\bar{\sigma}_2^2)}
\qquad .
\]
Again, the Hamilton-Jacobi equations for $H_I$ and $H_{II}$ must
be solved consecutively and the two pieces must be continuously
glued afterwards. The choice of sign combinations for ${\rm
s}_\lambda$, ${\rm s}_\mu$, ${\rm s}_\nu$ is opposite to that
giving the $K_3^{OD}(a,0)$ kinks.

These solutions asymptotically join the points $B^+$ and $B^-$
crossing the foci $F_3$. The kink form factors are:
\begin{eqnarray*}
\bar\phi_1(x)&=& \frac{(-1)^\alpha \sigma_3 \left(1+e^{2 \sqrt{2}
\bar\sigma_3^2  \bar{x}}\right)}{\sqrt{\sigma_3^2+ \bar\sigma_3^2
e^{-b}e^{2\sqrt{2} \bar x}+ e^{2\sqrt{2} \bar\sigma_3^2
 \bar{x}}} \sqrt{1+ \sigma_3^2
e^{2\sqrt{2}\bar\sigma_3^2  \bar
x}+ \bar\sigma_3^2 e^{b} e^{-2\sqrt{2}\sigma_3^2  \bar{x}}}} \\
\bar\phi_2(x)&=& 0\\
\bar\phi_3(x)&=&\frac{\sigma_3 \bar\sigma_3^2 \left(e^{2\sqrt{2}
 \bar{x}}-e^{b} \right) }{\sqrt{
 e^{2\sqrt{2}\sigma_3^2 \bar x}+ \sigma_3^2
e^{2\sqrt{2} \bar{x}}+\bar\sigma_3^2 e^{b}} \sqrt{e^{b}e^{2
\sqrt{2} \bar\sigma_3^2 \bar x}+ \sigma_3^2 e^{b}+\bar\sigma_3^2
e^{2\sqrt{2}\bar{x}}}} \label{eq:k4bb} \qquad .
\end{eqnarray*}
In Figure 14(a) a $K_6^{BB}$ kink is depicted with $\alpha=0$ that
connects the point $B_-$ with $B_+$. The structure of one member
of this family is similar to that found for the previous kinks.
From the energy density plot shown in Figure 14(b), one sees that
each member of the family consists of six basic particles: two
$K_1^{CD}$, two $K_1^{OB}$ and two $K_1^{BC}$ kinks. Only the
order differs (the two-body kink is on the right of the three-body
lump) and the two-body kink is made of a $K_1^{CD}$ and a
$K_1^{BC}$ solitary wave, instead of a $K_1^{OB}$ and a
$K_1^{BC}$. The energy, however, is the same:
\[
E(K_6^{BB})= E(K_1^{BO}) + E(K_3^{OD}(0,0))+ E(K_2^{DB})= 2
E(K_1^{CD}) + 2 E(K_1^{OB})+ 2 E(K_2^{BC})=\frac{1}{\sqrt{2}}
\qquad .
\]
\begin{figure}[hbt]
\centerline{\includegraphics[height=3cm]{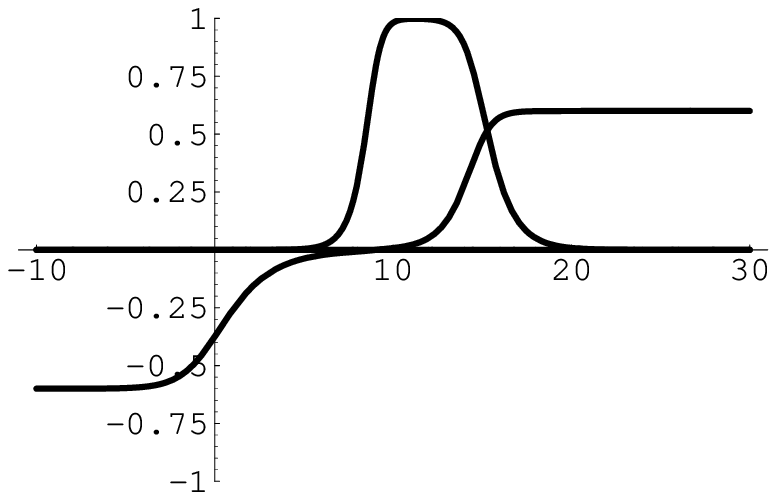}
\hspace{0.6cm}
\includegraphics[height=3cm]{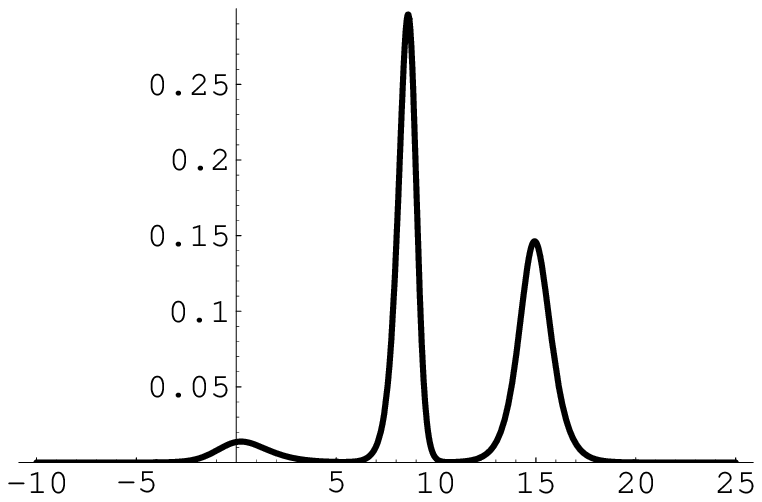} \hspace{0.6cm}
 \includegraphics[height=3cm]{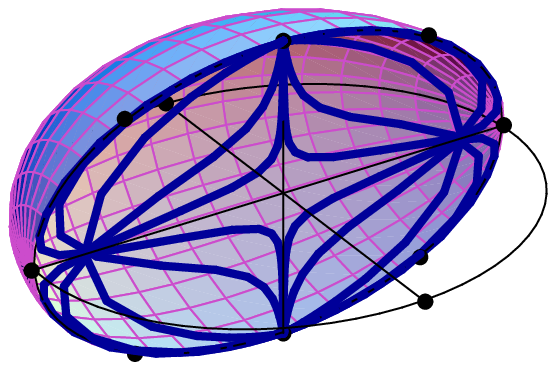}}
\caption{\small \textit{Kink form factor (a), energy density (b)
and orbits (c) for $K_6^{BB}$ kinks. }}
\end{figure}

\subsection{Seven-body solitary waves}

\textit{2. $K_7^{BC}(\gamma_2,\gamma_3)$:} Besides the generic
(depending on two parameters) $K_3^{OD}(a,b)$ kinks, there are
other generic solitary waves. Other combinations of signs in the
system of equations (\ref{eq:nu3}) provide new generic kinks
different from $K_3^{OD}(a,b)$. The kink orbits of these solutions
run between the $B$ and $C$ minima tracing a curve with three
well-defined stages: 1) First, the kink trajectory departs from
the point $C$ at $x=-\infty$ remaining in a octant of the
ellipsoid ${\mathbb E}$ interior until the particle reaches a
point of the characteristic hyperbola $h$. During this stage the
trajectory satisfies the system of equations (\ref{eq:nu3}) with
${\rm s}_\lambda ={\rm s}_\mu \neq {\rm s}_\nu $. 2) Then, the
kink orbit crosses to the next octant (from left to right or
viceversa) through this point and traces a curve that ends by
hitting a point in the characteristic ellipse $e_4$. In this
second stage the kink trajectory also complies with (\ref{eq:nu3})
but the sign combination is ${\rm s}_\lambda ={\rm s}_\nu \neq
{\rm s}_\mu$. 3) The last stage of the kink trajectory starts
again by changing the octant (from up to down or viceversa)
through the point in $e_4$. The kink orbit runs along a curve in
the latter octant until the point $B$ is reached. The final stage
satisfies the system (\ref{eq:nu3}) with ${\rm s}_\lambda \neq{\rm
s}_\mu = {\rm s}_\nu $, see Figure 15(c). The parameters
$\gamma_2$ and $\gamma_3$ respectively determine the points where
the kink orbit hits the characteristic hyperbola $h$ and ellipse
$e_4$. Orbits for which the relation $\gamma_2-\bar{\sigma}_3^2
\gamma_3$ between integration constants holds meet at a single
point on ellipse $e_4$. The same happens for trajectories that
comply with $\gamma_2-\bar{\sigma}_2^2 \gamma_3$ meeting at a
point on the hyperbola $h$. Thus, at each point in $h$ and $e_4$
an infinite number of kink trajectories meet. These characteristic
curves are focal lines.

These solutions are so complex that it is not possible to invert
the relationships (\ref{eq:nu3}) in order to obtain analytical
expressions in Cartesian coordinates for the kink form factors.
For this reason, we parametrize these solutions in terms of the
integration constants $\gamma_2$ and $\gamma_3$ that set the
orbits in the Hamilton-Jacobi expressions (\ref{eq:nu3}). The use,
however, of the implicit expressions in elliptic coordinates
allows Mathematica to plot field profiles and energy densities. It
should be mentioned that extreme care is necessary to guarantee
continuity of the trajectories and its derivatives when it
rebounds on the faces of the $P(0)$ cuboid. In Figure 15(a) the
form factor (field profile) of a $K_7^{BC}(\gamma_2,\gamma_3)$
kink is shown. In the Figure 15(b) we display the energy showing
seven basic lumps: two $K_1^{CD}$'s, two $K_1^{OB}$'s and three
$K_1^{BC}$'s, with the peculiarity that three different basic
particles always travel together. We remark that there exist no
net forces between particles in these configurations but that they
are highly unstable against any perturbation splitting the centers
of superposed particles. The total energy is:
\[
E(K_7^{BC})= 2 E(K_1^{CD}) + 2 E(K_1^{OB})+ 3
E(K_1^{BC})=\frac{1}{\sqrt{2}}+\frac{\bar\sigma_2^4-\bar\sigma_3^4}{2\sqrt{2}}
\]
\begin{figure}[hbt]
\centerline{\includegraphics[height=3cm]{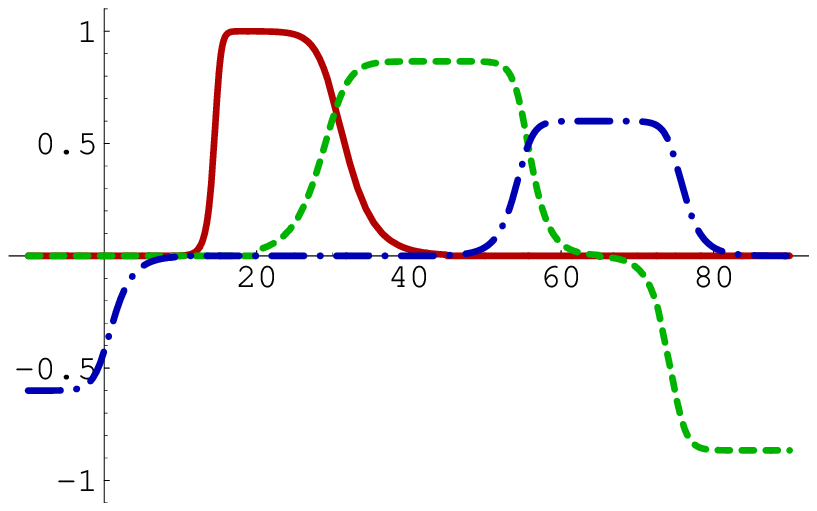}
\hspace{0.6cm}
\includegraphics[height=3cm]{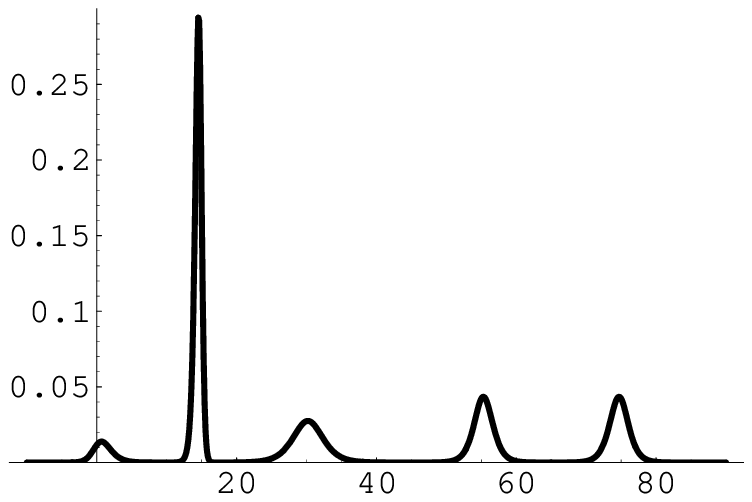} \hspace{0.6cm}
 \includegraphics[height=3cm]{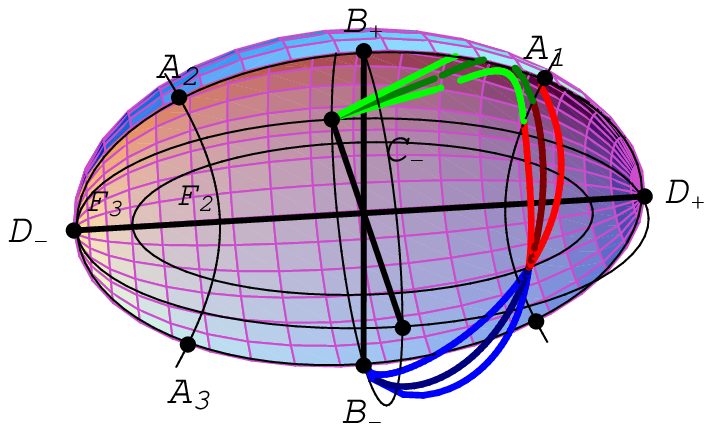}}
\caption{\small \textit{Kink form factor (a), energy density (b)
and orbits (c) for $K_7^{BC}$ kinks. }}
\end{figure}

A summary of all the kink energies is offered in the following
table:

\begin{tabular}{|l||c|} \hline
\textbf{Type of Kink} & \textbf{Energy} \\ \hline\hline
\textit{Basic Kinks} & \rule{0cm}{0.6cm}
$E(K_1^{OB})=\frac{\bar\sigma_3^4}{2\sqrt{2}}$ \hspace{1cm}
$E(K_1^{CD})=\frac{\sigma_2^2(2-\sigma_2^2))}{2\sqrt{2}}$
\hspace{1cm}
$E(K_1^{BC})=\frac{\bar\sigma_2^4-\bar\sigma_3^4}{2\sqrt{2}}$ \\
\hline \textit{Two-Body Kinks} & \rule{0cm}{0.6cm}
$E[K_2^{OC}(b)]=\frac{\bar\sigma_2^4}{2\sqrt{2}}$ \hspace{1cm}
$E[K_2^{BD}(b)]=\frac{\sigma_3^2(2-\sigma_3^2))}{2\sqrt{2}}$ \\
\hline \textit{Three-Body Kinks} & \rule{0cm}{0.6cm}
$E[K_3^{OD}(a,b)]=\frac{1}{2\sqrt{2}}$ \\
\hline \textit{Four-Body Kinks} & \rule{0cm}{0.6cm}
$E[K_4^{BB}(b)]=\frac{\bar\sigma_2^4}{\sqrt{2}}$ \hspace{1cm}
$E[K_4^{CC}(b)]=\frac{\sigma_3^2(2-\sigma_3^2))}{\sqrt{2}}$ \\
\hline \textit{Six-Body Kinks} & \rule{0cm}{0.6cm}
$E[K_6^{CC}(b)]=\frac{1}{\sqrt{2}}$ \hspace{1cm}
$E[K_6^{BB}(b)]=\frac{1}{\sqrt{2}}$ \\
\hline \textit{Seven-Body Kinks} & \rule{0cm}{0.6cm}
$E[K_7^{BC}(a,b)]=\frac{1}{\sqrt{2}}+\frac{\bar\sigma_2^4-\bar\sigma_3^4}{2\sqrt{2}}$
 \\ \hline
\end{tabular}

\end{document}